\documentclass[journal,10pt]{IEEEtran}
\IEEEoverridecommandlockouts
\usepackage{graphicx}
\usepackage{caption}
\usepackage{enumerate}
\usepackage{epsfig}
\usepackage{epstopdf}
\usepackage{svg}
\usepackage{cite}
\usepackage{subfigure}
\usepackage{amsmath,amssymb,amsfonts}
\usepackage{algorithmic}
\usepackage{tabularx}
\usepackage{textcomp}
\usepackage{xcolor}
\usepackage{stfloats}
\usepackage{multirow}
\usepackage{url}
\usepackage[ruled]{algorithm2e}
\usepackage{hyperref}
\usepackage{etoolbox,xstring,mfirstuc,textcase}
\usepackage{bm}
\usepackage{amsmath}
\usepackage{amsthm}
\usepackage{stackengine}
\usepackage{cuted}
\usepackage{makecell}
\usepackage{booktabs}
\usepackage{threeparttable}
\usepackage[skip=3pt]{caption}
\usepackage[explicit]{titlesec}

\newtheorem{remark}{Remark}
\def\BibTeX{{\rm B\kern-.05em{\sc i\kern-.025em b}\kern-.08em
    T\kern-.1667em\lower.7ex\hbox{E}\kern-.125emX}}

\begin{document}

\title{A~Fully~Wave-Domain~Wideband~MU-MIMO~OFDM Transmitter via~Stacked~Intelligent~Metasurfaces} 
	\author{Zheao Li, \IEEEmembership{Graduate Student Member, IEEE}, Jiancheng An, \IEEEmembership{Member, IEEE}, and Chau Yuen, \IEEEmembership{Fellow, IEEE}
\vspace{-1.3 cm}
\thanks{Z. Li, J. An, and C. Yuen are with the School of Electrical and Electronic Engineering, Nanyang Technological University, 639798, Singapore (email: zheao001@e.ntu.edu.sg, \{jiancheng.an, chau.yuen\}@ntu.edu.sg). (\emph{Corresponding author: Chau Yuen}.)}
}

\markboth{DRAFT}%
{Shell \MakeLowercase{\textit{et al.}}: A Sample Article Using IEEEtran.cls for IEEE Journals}

\maketitle
\begin{abstract}
This paper proposes an advanced realization principle for wideband multiuser multiple-input multiple-output orthogonal frequency-division multiplexing (MU-MIMO OFDM) transmitters, where the conventional transmitter-side baseband chain is physically synthesized in the wave domain. For design and optimization purposes, this fully wave-domain wideband MU-MIMO OFDM transmitter implemented by a cascaded SIM structure is functionally partitioned into two cascaded SIM blocks. The first block, denoted as SIM$_1$, integrates symbol loading and channel-adaptive MU-MIMO precoding updated at the channel-coherence timescale, mapping the user streams to a virtual port-subcarrier representation. The second block, denoted as SIM$_2$, acts as an offline-configured sampling-rate modulator that materializes the inverse discrete Fourier transform (IDFT) and cyclic prefix (CP) insertion directly in the wave domain. This baseband-free architecture establishes a virtual-to-physical transition from information bits to radiated CP-extended OFDM waveforms. To account for practical nonidealities, SIM$_2$ is optimized to fit the ideal multi-port CP-OFDM operator, and its residual response is mapped into an effective coupling matrix. Then, SIM$_1$ is optimized in a communication-oriented manner by jointly adapting discrete phase shifts and stream-subcarrier power loading to maximize the sum spectral efficiency. Results demonstrate the convergence, architecture trade-off, wave-domain OFDM materialization accuracy, and competitive performance of the proposed baseband-free transmitter.
\end{abstract}
\vspace{-0.3 cm}
\section{Introduction}

M{odern} high-speed wireless communication systems are increasingly expected to support massive data traffic, high spectral efficiency, low latency, and reliable user connectivity \cite{vision1}. To meet these requirements, multiple-input multiple-output (MIMO) transmission and multicarrier modulation have become two foundational technologies in contemporary wireless networks. In particular, orthogonal frequency-division multiplexing (OFDM) is widely adopted to combat frequency selectivity in wideband channels, while multiuser MIMO (MU-MIMO) enables spatial multiplexing and interference management across multiple users  \cite{OFDM, OFDM2, OFDM3}. By combining spatial-domain multiplexing with frequency-domain parallel transmission, MU-MIMO OFDM has become a fundamental air-interface technology for high-throughput wireless systems. However, in high-frequency, wideband, and large-array regimes, its transmitter-side implementation becomes increasingly hardware-intensive. A conventional fully-digital base station (BS) needs to perform stream loading, subcarrier-wise multiuser precoding, inverse discrete Fourier transform (IDFT), cyclic prefix (CP) insertion, and high-speed digital-to-analog conversion (DAC) before radiation \cite{FD}. As the numbers of antennas, users, and subcarriers increase, this architecture incurs substantial baseband computing complexity, high radio-frequency (RF) chain cost, stringent DAC requirements, and considerable power consumption. These burdens become particularly critical for millimeter-wave and wideband large-array systems, where the transmitter must process many space-frequency coefficients within each OFDM block.

To alleviate these burdens, hybrid analog and digital precoding has been widely investigated for millimeter-wave and massive MIMO systems \cite{HY1, HY2, HY3}. By exploiting a low-dimensional digital precoder with an analog radio-frequency network, hybrid architectures can reduce the number of RF chains while retaining part of the spatial multiplexing gain. Various wideband extensions, including subarray-based structures, frequency-selective analog networks, and beam-squint-aware designs \cite{new1, new2, new3}, have also been developed to improve broadband performance. Nevertheless, these architectures still preserve the fundamental digital baseband pipeline. In particular, data streams are digitally precoded across subcarriers, transformed into time-domain OFDM samples through IDFT, and appended with a CP before upconversion and transmission. Therefore, hybrid precoding mainly reduces the spatial-domain precoding burden, while the high-speed digital OFDM waveform-generation chain remains largely unchanged.

Stacked intelligent metasurfaces (SIMs) offer an alternative hardware paradigm for implementing signal transformations through controllable electromagnetic (EM) propagation \cite{SIM0}. Early efforts demonstrated that cascaded diffractive layers can implement trainable EM transformations, drawing an analogy between multilayer metasurfaces and optical neural networks \cite{Lin_Science_2018_D2NN,Liu_NatureElec_2022_ProgrammableD2NN}. Building upon this idea, SIMs have shown strong potential to implement controllable diffractive transformations for wireless communications \cite{survey2}. Specifically, SIM-enabled holographic MIMO systems were investigated in \cite{An_JSAC_2023_HolographicSIM,Papazafeiropoulos_TWC_2024_HolographicSIM}, showing that stacked metasurface layers can provide additional spatial degrees of freedom (DoF) for efficient MIMO transmission. MU downlink beamforming and radiation-pattern control using SIM were studied in \cite{An_ICC_2023_MUBeamforming,Hassan_OJCOMS_2024_Beamforming,An_TWC_2025_MUBeamforming}, where the SIM was mainly used for spatial precoding or inter-user interference suppression in the wave domain \cite{Nadeem_WCNC_2024_ChannelEst,Yao_WCL_2024_ChannelEst}. The potential of SIM has also been extended to near-field communications \cite{Papazafeiropoulos_WCL_2024_NearField}, direction-of-arrival estimation \cite{An_JSAC_2024_DOA}, satellite communications \cite{Lin_WCL_2024_LEO}, secure transmission \cite{Niu_TIFS_2024_SecureSIM}, semantic communications \cite{Huang_WCL_2025_SemanticSIM}, cell-free networks \cite{Li_TCOM_2024_CellFreeSIM,Shi_TWC_2025_CellFreeSIM}, and integrated sensing and communications \cite{Niu_WCL_2024_ISACSIM}. With its configured transmission coefficients acting as trainable parameters, SIM has also been investigated as a physical platform for over-the-air computation \cite{survey1, NN}. 

Although these studies have firmly established SIM as a promising wave-domain computing architecture, most existing designs still treat SIM as an auxiliary spatial processing module, such as a beamformer, channel shaper, or spatial processor \cite{N1}. The corresponding models are typically narrowband or mainly focus on frequency-flat spatial transformations \cite{NBF}. Recent wideband SIM studies have further shown that multilayer metasurfaces can enhance spatial multiplexing and compensate frequency-selective fading over multiple subcarriers \cite{Li1, Li2, Li3, Li4}. However, even in these wideband designs, the conventional OFDM waveform-generation chain, including IDFT and CP insertion, is still performed in the digital baseband. Therefore, existing SIM-based wideband systems do not yet physically synthesize the complete transmitter-side MU-MIMO OFDM processing chain. This gap motivates a more fundamental question: can the transmitter itself be decomposed and realized in the wave domain?

To answer this question, this paper moves beyond the SIM-assisted spatial beamforming component and proposes a cascaded SIM transmitter architecture for physically synthesizing the transmitter-side processing chain itself. For clarity of design, modeling, and optimization, the overall multilayer structure is functionally divided into two consecutive SIM blocks. Specifically, the first block, denoted as SIM$_1$, performs integrated symbol loading and channel-adaptive wave-domain MU-MIMO precoding. It is updated at the channel-coherence timescale and maps the loaded user streams into a virtual space-frequency representation. The second block, denoted as SIM$_2$, acts as a fast spatio-temporal wave-domain modulator. It is optimized offline to materialize the IDFT and CP operation and is reused during online transmission. This two-SIM decomposition naturally matches the distinct roles of spatial precoding and temporal OFDM modulation, thereby enabling a baseband-free wideband MU-MIMO OFDM transmitter. However, practical SIM$_2$ can only approximate the ideal CP-OFDM modulation operator due to finite metasurface dimensions and discrete phase constraints, which would result in residual cross-subcarrier coupling after receiver-side OFDM demodulation. To address this issue, we first formulate the offline SIM$_2$ synthesis as an operator-fitting problem with respect to the ideal multi-port IDFT-and-CP operator. The fitted SIM$_2$ response is then mapped into an effective coupling matrix after receiver-side CP removal and discrete Fourier transform, where this matrix explicitly captures the residual inter-carrier interference (ICI) induced by imperfect SIM$_2$ materialization. Under this fixed nonideal coupling, SIM$_1$ is further optimized in a communication-oriented manner. Instead of merely fitting a prescribed digital zero-forcing (ZF) precoder, SIM$_1$ jointly adapts layer-wise discrete phase shifts and stream-subcarrier power-loading coefficients to maximize the end-to-end CP-aware sum spectral efficiency.

The main contributions and innovations of this research are described as follows.

\begin{itemize}
    \item We propose a wave-domain architecture for wideband MU-MIMO OFDM transmission via a cascaded SIM structure. SIM$_1$ performs channel-adaptive symbol loading and MU-MIMO precoding at the channel-coherence timescale, while SIM$_2$ implements channel-independent CP-OFDM waveform generation through a preconfigured spatio-temporal modulation sequence. This cascaded architecture shifts the high-dimensional precoding and OFDM waveform-generation operations from digital baseband processing to programmable EM propagation.

    \item We formulate the offline SIM$_2$ design as an operator-fitting problem with respect to the ideal multi-port IDFT-and-CP operator. The fitted SIM$_2$ response is further mapped into an effective coupling matrix, which explicitly characterizes the residual ICI caused by nonideal wave-domain OFDM materialization.

    \item We formulate a communication-oriented SIM$_1$ optimization problem under the fixed fitted-SIM$_2$ coupling matrix. Unlike pure digital-ZF fitting, the proposed formulation jointly optimizes the discrete phase shifts and stream-subcarrier power-loading coefficients, thereby allowing SIM$_1$ to adapt to both the wireless channel and the residual coupling of the compiled OFDM modulator.

    \item We design a two-stage optimization framework. SIM$_2$ is optimized offline using continuous-relaxation-aided discrete refinement, while SIM$_1$ is optimized online using fractional programming (FP), Karush-Kuhn-Tucker (KKT) based power loading, and layer-wise projected-gradient with exact block-coordinate phase (E-BCD) refinement. Numerical results validate the convergence behavior, architecture trade-off, wave-domain CP-OFDM materialization accuracy, and end-to-end spectral-efficiency performance of the proposed transmitter.
\end{itemize}

The remainder of this paper is organized as follows. Section~II is the system model. Section~III formulates the SIM$_2$ OFDM materialization problem and the SIM$_1$ communication-oriented design problem. Section~IV develops the proposed two-stage optimization framework. Section~V provides numerical results, and Section~VI concludes this paper.

\textbf{Notations}: \(\Re\{\cdot\}\) and \(\Im\{\cdot\}\) denote the real and imaginary parts of a complex scalar, respectively; \(\angle(\cdot)\) denotes the phase angle; \(\odot\) denotes the Hadamard product; \(\circ\) denotes the Khatri--Rao product; and \(\succeq \mathbf 0\) denotes positive semidefiniteness.
\vspace{-0.2 cm}
\section{System Model}

In this section, we present the system model of the proposed baseband-free wave-domain transmitter, where symbol loading, MU-MIMO precoding, and OFDM modulation are redistributed from digital baseband to the cascaded SIMs.
\vspace{-0.2 cm}
\subsection{Architecture Overview and Design Rationale}

\begin{figure*}
	\centerline{\includegraphics[width=0.9\textwidth]{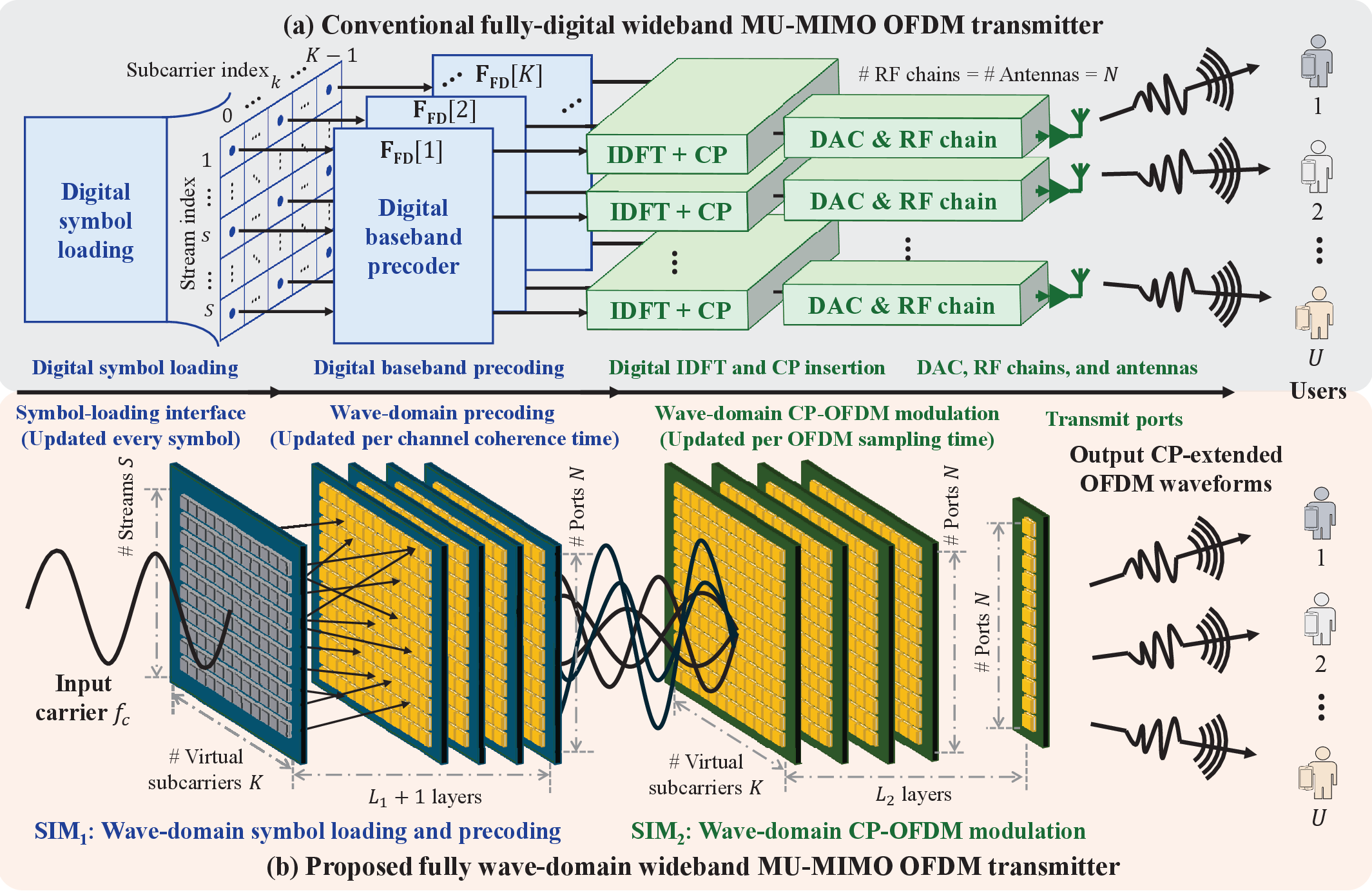}}
	\caption{Comparison of (a) the conventional fully-digital wideband MU-MIMO OFDM transmitter and (b) the proposed baseband-free fully wave-domain wideband MU-MIMO OFDM transmitter.}
	\label{Fig0}
\vspace{-0.5 cm}
\end{figure*}

We consider a downlink MU-MIMO OFDM system, where a BS serves $U$ single-antenna users over $K$ subcarriers by our baseband-free wave-domain transmitter. Let $S$ denote the number of transmitted data streams, and we focus on the one-stream-per-user setting $U\!=\!S\!<\!N$, where $N$ is the number of physical transmit ports after the proposed cascaded metasurface structure. Let $\mathcal{K}\triangleq\{0,1,\ldots,K-1\}$ denote the subcarrier index set, and let $\mathbf{s}[k]\in\mathbb{C}^{S\times 1}$ denote the stream vector transmitted on subcarrier $k$, with $\mathbb{E}[\mathbf{s}[k]\mathbf{s}^{H}[k]]=\mathbf I_S,~\forall k\in\mathcal{K}$.

As shown in Fig.~\ref{Fig0}(a), a conventional fully-digital wideband transmitter first performs symbol loading and subcarrier-wise MU-MIMO precoding in the digital baseband. Its frequency-domain transmit vector on subcarrier $k$ can be written as
\begin{equation}
    \mathbf{x}^{\mathrm{FD}}[k]
    =
    \mathbf{F}_{\mathrm{FD}}[k]\boldsymbol{\Lambda}_{\rho}^{\mathrm{FD}}[k]\mathbf{s}[k]
    \in\mathbb{C}^{N\times 1},
    \label{eq:conv_fully_digital}
\end{equation}
where $\mathbf{F}_{\mathrm{FD}}[k]\in\mathbb{C}^{N\times S}$ is the fully-digital precoder and $\boldsymbol{\Lambda}_{\rho}^{\mathrm{FD}}[k]\in\mathbb{R}^{S\times S}_{+}$ is the stream-loading matrix. The $K$ frequency-domain vectors are converted into $N$ CP-extended time-domain signals by per-antenna IDFT and CP insertion, followed by high-resolution DACs, RF chains, and radiation.

In contrast, Fig. 1(b) shows the proposed baseband-free fully wave-domain transmitter. For design clarity, this structure is functionally partitioned into two consecutive SIM blocks:
\begin{itemize}
    \item \textbf{SIM$_1$}: performs integrated symbol loading, stream-subcarrier power loading, and channel-adaptive wave-domain MU-MIMO precoding;
    \item \textbf{SIM$_2$}: performs wave-domain CP-OFDM materialization as the physical counterpart of IDFT and CP insertion.
\end{itemize}
In practice the two blocks can be implemented as consecutive regions of a single integrated multilayer metasurface device.

A defining feature of this baseband-free design is its intrinsic timescale separation, as summarized in Table \ref{table:comparison}. Specifically, SIM$_1$ performs channel-adaptive wave-domain MU-MIMO precoding at the slow channel-coherence timescale, whereas SIM$_2$ executes channel-independent CP-OFDM waveform materialization at the fast sampling timescale and can therefore be optimized offline.

\begin{table}[t]
\centering
\caption{\centering{\protect\\{\textsc{Comparison of Functional Roles of two SIMs.}}}}\
\label{selection} 
	\setlength{\tabcolsep}{1pt} 
	\renewcommand\arraystretch{1.8} 
\label{table:comparison}
\begin{tabular}{|m{2.5cm}<{\centering}|m{3.1cm}<{\centering}|m{3cm}<{\centering}|}
\hline
\textbf{Characteristic} & \textbf{SIM$_1$ (Precoder)} & \textbf{SIM$_2$ (Modulator)} \\ \hline 
Operation domain & Space-frequency domain / slow-timescale & Time domain / fast-timescale \\ \hline
Update frequency & Channel coherence & Baseband sampling \\ \hline
Input excitation & Stream-subcarrier symbols & Virtual port-subcarrier signals \\ \hline
Primary function & MU-MIMO precoding & CP-OFDM modulation \\ \hline
Design strategy & Online communication-oriented & Offline operator-fitting and reused \\ \hline
\end{tabular}
\vspace{-0.5 cm}
\end{table}

Throughout Section II, we focus on one generic CP-extended OFDM block. The block index is omitted since a sufficiently long CP perfectly removes inter-block interference, allowing independent analysis. For one OFDM block, we define the stacked stream vector
\begin{equation}
    \mathbf{s}_{\mathrm v}
    \triangleq
    \begin{bmatrix}
        \mathbf{s}^{\top}[0],~
        \mathbf{s}^{\top}[1],~
        \ldots,~
        \mathbf{s}^{\top}[K-1]
    \end{bmatrix}^{\top}
    \in\mathbb{C}^{SK\times 1},
    \label{eq:sv_def_final}
\end{equation}
which serves as the dynamic data input of the wave-domain transmitter. In the sequel, we detail how $\mathbf{s}_{\mathrm v}$ is first loaded and precoded by SIM$_1$, then transformed by SIM$_2$ into a radiated CP-extended OFDM waveform, and finally interpreted at the receiver side as a physical wideband OFDM transmission.

\begin{remark}
\textnormal{In this paper, the subcarrier index in $\mathbf{s}_{\mathrm v}$ is a logical or virtual OFDM-mode index rather than a physical frequency index. Before SIM$_2$ performs the time-domain waveform materialization, the entries associated with different $k\in\mathcal{K}$ merely label the intended modes to be processed by the wave-domain operator. The physical OFDM subcarriers emerge only after SIM$_2$ converts this virtual space-frequency representation into a temporal waveform and the receivers subsequently perform standard CP removal and DFT.}
\end{remark}
\vspace{-0.2 cm}
\subsection{SIM$_1$-Based Symbol Loading and MU-MIMO Precoding}

SIM$_1$ is modeled as a unified $(L_1+1)$-layer wave-domain processor that jointly realizes integrated symbol loading and MU-MIMO precoding. Specifically, the $0$-th layer of SIM$_1$ is updated at the OFDM-symbol timescale, while the subsequent $L_1$ layers are updated at the channel-coherence-block timescale for channel-adaptive precoding.

According to \cite{MultiMod} and \cite{MultiMod2}, the metasurface layer 0 of SIM$_1$ can load the symbol block $\{{\mathbf{s}}[k]\}_{k\in\mathcal{K}}$ onto the common carrier $f_c$ , and inject the resulting wave-domain excitation into the subsequent cascades. To incorporate integrated power loading at the symbol-loading interface, let $\rho_{u,k}\ge 0$ denote the power-loading coefficient assigned to user stream $u$ on the $k$-th virtual subcarrier, and define $\boldsymbol{\Lambda}_{\rho}[k] \triangleq
\operatorname{diag}\left(\sqrt{\rho_{1,k}},\ldots,\sqrt{\rho_{U,k}}\right) \in \mathbb{R}^{S\times S}_{+}.$ Stacking all virtual subcarriers gives
\begin{align}
    \tilde{\mathbf{s}}_{ v}
    &=
    \operatorname{blkdiag}
    \left(
        \boldsymbol{\Lambda}_{\rho}[0],\ldots,\boldsymbol{\Lambda}_{\rho}[K-1]
    \right)\mathbf s_v,
    \label{eq:loaded_sv_SIM1_final}\\
    &=\begin{bmatrix}
        \tilde{\mathbf{s}}^\top[0],~
        \tilde{\mathbf{s}}^\top[1],~
        \cdots,~
        \tilde{\mathbf{s}}^\top[K-1]
    \end{bmatrix}^\top
    \in\mathbb{C}^{SK\times 1}.
\end{align}

To describe the forward propagation of SIM$_1$, let $d_l$ denote the port dimension of the $l$-th metasurface layer:
\begin{equation}
    d_l
    =
    \begin{cases}
        SK, & l=0,\\
        M_1, & l=1,\ldots,L_1-1,\\
        NK, & l=L_1.
    \end{cases}
    \label{eq:dl_SIM1_final2}
\end{equation}
where the layer $0$ serves as a symbol-loading interface with $SK$ ports, the first $L_1-1$ layers operate on the hidden $M_1$-dimensional space with $M_1=M_{x,1}\times K$, while the final layer is an $NK$-dimensional virtual-port interface.

For $l\in\mathcal{L}_1\triangleq\{1,\ldots,L_1\}$, let $\mathbf{W}^{l}\in\mathbb{C}^{d_l\times d_{l-1}}$ denote the propagation matrix from the $(l\!-\!1)$-th layer to the $l$-th layer of SIM$_1$. In particular, $\mathbf{W}^{1}\in\mathbb{C}^{M_1\times SK}$ describes the propagation from the $0$-th symbol-loading layer to the first layer, $\mathbf{W}^{l}\in\mathbb{C}^{M_1\times M_1}$, $l=2,\ldots,L_1\!-\!1$, describe the propagation between adjacent hidden computational layers, and $\mathbf{W}^{L_1}\in\mathbb{C}^{NK\times M_1}$ denotes the propagation matrix spanning from the last $M_1$-dimensional hidden representation to the $NK$-port $L_1$-th layer of SIM$_1$. For notational convenience, define the port-index set of the $l$-th layer of SIM$_1$ as $\mathcal{D}_l
    \triangleq
    \{1,2,\ldots,d_l\}$ for $l\in\mathcal{L}_1$.
Accordingly, for $\mathbf{W}^{l}\in\mathbb{C}^{d_l\times d_{l-1}}$, its entry $[\mathbf{W}^{l}]_{m,m'}$ is indexed by $m\in\mathcal{D}_l$ and $m'\in\mathcal{D}_{l-1}$:
\begin{equation}
    [\mathbf{W}^{l}]_{m,m'}
    =
    \frac{A_1 \delta_1}{(r_{m,m'}^{l})^2}
    \left(
        \frac{1}{2\pi r_{m,m'}^{l}}
        -
        j\frac{f_c}{c}
    \right)
    e^{j2\pi f_c r_{m,m'}^{l}/c},
    \label{eq:W_RS_SIM1_final2}
\end{equation}
where $A_1$ is the meta-atom area of SIM$_1$, $\delta_1$ is the inter-plane spacing, $r_{m,m'}^{l}$ is the propagation distance between the corresponding meta-atoms, and $c$ is the speed of light.

The phase shifts of the $l$-th layer are represented by
\begin{equation}
    \mathbf{\Phi}^{l}
    =
    \operatorname{diag}
    \big(
        e^{j\theta^{l}_{1}},
        \ldots,
        e^{j\theta^{l}_{d_l}}
    \big)
    \in\mathbb{C}^{d_l\times d_l},
    \quad
    l\in\mathcal{L}_1,
    \label{eq:Phi_SIM1_final2}
\end{equation}
where $\theta_m^l$ denotes the phase shift applied by the $m$-th meta-atom on the $l$-th layer. Collecting all layer-wise phase-shifts matrices into $\mathbf{\Phi}
    \triangleq
    \left\{
        \mathbf{\Phi}^{l}
    \right\}_{l=1}^{L_1}$, the operator of SIM$_1$ is
\begin{equation}
    \mathbf{P}(\mathbf{\Phi})
    =
    \mathbf{\Phi}^{L_1}
    \mathbf{W}^{L_1}
    \cdots
    \mathbf{\Phi}^{1}
    \mathbf{W}^{1}
    \in\mathbb{C}^{NK\times SK}.
    \label{eq:P_Phi_SIM1_final2}
\end{equation}

To maintain the separability of the virtual OFDM modes before SIM$_2$-based waveform materialization, we impose a block-diagonal architectural constraint on SIM$_1$. Specifically, the SIM$_1$ ports are grouped into $K$ virtual subcarrier branches, and cross-branch mappings are structurally suppressed by design. Therefore, the feasible SIM$_1$ operator is restricted to
\begin{equation}
    \mathbf{P}(\mathbf{\Phi})
    =
    \operatorname{blkdiag}\!\big(
        \mathbf{P}[0],\mathbf{P}[1],\ldots,\mathbf{P}[K-1]
    \big),
    \label{eq:P_blkdiag_final}
\end{equation}
where $\mathbf{P}[k]\in\mathbb{C}^{N\times S}$ denotes the effective precoding block associated with the $k$-th virtual subcarrier.

The virtual space-frequency output of SIM$_1$ is given by
\begin{align}
    \mathbf{x}_{\mathrm v}
    &=
    \mathbf P(\mathbf \Phi)
    \operatorname{blkdiag}
    \left(
        \boldsymbol{\Lambda}_{\rho}[0],\ldots,\boldsymbol{\Lambda}_{\rho}[K-1]
    \right)
    \mathbf s_v
    \label{eq:xv_sim1_final}
    \\
    &\triangleq
    \begin{bmatrix}
        \mathbf{x}^\top[0],~\cdots,~\mathbf{x}^\top[K{-}1]
    \end{bmatrix}^\top
    \in \mathbb{C}^{NK\times 1},
\end{align}
and for the $k$-th virtual subcarrier $\mathbf x[k] \in\mathbb{C}^{N\times 1}$,
\begin{align}
    \mathbf x[k]
    =
    \mathbf P[k]\boldsymbol{\Lambda}_{\rho}[k]\mathbf s[k]
    =
    \sum_{u=1}^{U}
    \sqrt{\rho_{u,k}}\mathbf p_u[k]s_u[k],
\end{align}
where $\mathbf p_u[k]$ denotes the $u$-th column of $\mathbf P[k]$, corresponding to the precoder for user stream $u$ on the $k$-th virtual subcarrier.
\vspace{-0.8 cm}
\subsection{SIM$_2$-Based Wave-Domain OFDM Modulation}

Unlike SIM$_1$, SIM$_2$ is modeled as a high-speed spatio-temporal modulator and reused during online transmission. Its role is to materialize the mathematical IDFT and CP-insertion operations directly in the wave domain by transforming the $NK$-dimensional port-subcarrier vector generated by SIM$_1$.

Let $T\triangleq K+N_{\mathrm{CP}}$ denote the OFDM block length after CP insertion, where $N_{\mathrm{CP}}$ is the CP length. For the layers of SIM$_2$, indexed by $\ell\in\mathcal{L}_2\triangleq\{1,\ldots,L_2\}$, define the layer dimension
\begin{equation}
    \bar d_{\ell}
    =
    \begin{cases}
        NK, & \ell=1,\\
        M_2, & \ell=2,\ldots,L_2-1,\\
        N, & \ell=L_2.
    \end{cases}
    \label{eq:dl_SIM2_B}
\end{equation}
where the first layer of SIM$_2$ is an $NK$-port input layer, the intermediate layers operate on the hidden $M_2$-dimensional space with $M_2=M_{x,2}\times K$, and the final layer is directly represented on the $N$ physical transmit ports. $\bar d_{0}=NK$ represents the virtual $0$-th interface inherited from SIM$_1$.

For $\ell\in\mathcal{L}_2$, let $\mathbf{U}^{\ell}\in\mathbb{C}^{\bar d_{\ell}\times \bar d_{\ell-1}}$ denote the propagation matrix from the $(\ell-1)$-th layer to the $\ell$-th layer of SIM$_2$. $\mathbf{U}^{1}\in\mathbb{C}^{NK\times NK}$ describes the inter-module propagation from the $NK$-dimensional virtual output interface of SIM$_1$ to the independently programmable $NK$-port input layer of SIM$_2$, $\mathbf{U}^{2}\in\mathbb{C}^{M_2\times NK}$ maps the first $NK$-port programmable layer to the second hidden computational layer, $\mathbf{U}^{\ell}\in\mathbb{C}^{M_2\times M_2}$ with $\ell=3,\ldots,L_2-1$ represent the propagation between adjacent hidden computational layers, and $\mathbf{U}^{L_2}\in\mathbb{C}^{N\times M_2}$ maps the last hidden layer to the final $N$-port output layer of SIM$_2$.

For $\ell=0,1,\ldots,L_2$, define the port-index set of the $\ell$-th layer as $\bar{\mathcal{D}}_{\ell} \triangleq \{1,2,\ldots,\bar d_{\ell}\}$. Hence, for $\mathbf{U}^{\ell}\in\mathbb{C}^{\bar d_{\ell}\times \bar d_{\ell-1}}$, its entry $[\mathbf{U}^{\ell}]_{\mu,\mu'}$ is indexed by $\mu\in\bar{\mathcal{D}}_{\ell}$ and $\mu'\in\bar{\mathcal{D}}_{\ell-1}$:
\begin{equation}
    [\mathbf{U}^{\ell}]_{\mu,\mu'}
    =
    \frac{A_2 \delta_2}{(\tilde r_{\mu,\mu'}^{\ell})^2}
    \left(
        \frac{1}{2\pi \tilde r_{\mu,\mu'}^{\ell}}
        -
        j\frac{f_c}{c}
    \right)
    e^{j2\pi f_c \tilde r_{\mu,\mu'}^{\ell}/c},
    \label{eq:U_RS_SIM2_final2}
\end{equation}
where $A_2$ is the meta-atom area of SIM$_2$, $\delta_2$ is the inter-plane spacing, and $\tilde r_{\mu,\mu'}^{\ell}$ is the corresponding propagation distance.

At each time index $t\in\mathcal{T}\triangleq\{0,1,\ldots,T-1\}$, the phase shifts of the $\ell$-th programmable layer are represented by
\begin{equation}
    \mathbf{\Psi}^{\ell}[t]
    =
    \operatorname{diag}
    \big(
        e^{j\zeta_{1}^{\ell}[t]},
        \ldots,
        e^{j\zeta_{\bar d_{\ell}}^{\ell}[t]}
    \big)
    \in\mathbb{C}^{\bar d_{\ell}\times \bar d_{\ell}},
    \quad
    \ell\in\mathcal{L}_2,
    \label{eq:Psi_SIM2_final2}
\end{equation}
where $\zeta_{\mu}^{\ell}[t]$ denotes the time-varying phase shift applied by the $\mu$-th meta-atom on the $\ell$-th layer.

The instantaneous operator of SIM$_2$ at time index $t$ is
\begin{equation}
    \mathbf{Q}({\mathbf{\Psi}}[t])
    =
    \mathbf{\Psi}^{L_2}[t]
    \mathbf{U}^{L_2}
    \cdots
    \mathbf{\Psi}^{1}[t]
    \mathbf{U}^{1}
    \in\mathbb{C}^{N\times NK}.
    \label{eq:Qt_SIM2_B}
\end{equation}

Throughout one OFDM block of duration $T\times T_s$, where $T_s$ is the sampling interval, the virtual port-subcarrier excitation $\mathbf{x}_{\mathrm v}$ generated by SIM$_1$ is held constant, while SIM$_2$ follows a prescribed sequence of \(T\) phase patterns over one CP-extended OFDM block, with the intended switching interval matched to the sampling period \(T_s\) \cite{stc}. Consequently, the instantaneous radiated signal vector at time index $t$ is synthesized as
\begin{equation}
    \mathbf{x}_{\mathrm t}[t] = \mathbf{Q}({\mathbf{\Psi}}[t]) \mathbf{x}_{\mathrm v} \in \mathbb{C}^{N\times 1}, \quad t \in \mathcal{T}.
\end{equation}

Collecting all phase matrices yields $
    \bar{\mathbf{\Psi}}
    \triangleq
    \left\{
        \mathbf{\Psi}^{\ell}[t]
    \right\}_{\ell\in\mathcal{L}_2,\,t\in\mathcal{T}}$. Stacking the operators over all $T$ time indices gives the overall time-domain modulation operator
\begin{equation}
    \mathbf{Q}(\bar{\mathbf{\Psi}})
    \triangleq
    \begin{bmatrix}
        \mathbf{Q}^\top({\mathbf{\Psi}}[0]),
        \cdots,
        \mathbf{Q}^\top({\mathbf{\Psi}}[T-1])
    \end{bmatrix}^\top
    \in\mathbb{C}^{NT\times NK}.
    \label{eq:Qall_SIM2_B}
\end{equation}
and the total radiated CP-extended OFDM waveform is then
\begin{align}
    \mathbf{x}_{\mathrm{time}}
    &=
    \mathbf{Q}(\bar{\mathbf{\Psi}})\mathbf{x}_{\mathrm v}\\
   & \triangleq
    \begin{bmatrix}
        \mathbf{x}_{\mathrm t}^{\top}[0],~
        \mathbf{x}_{\mathrm t}^{\top}[1],~
        \ldots,~
        \mathbf{x}_{\mathrm t}^{\top}[T-1]
    \end{bmatrix}^{\top}
    \in\mathbb{C}^{NT\times 1},
    \label{eq:xtime_SIM2_B}
\end{align}
where $\mathbf{x}_{\mathrm t}[t]\in\mathbb{C}^{N\times 1}$ denotes the transmitted $N$-dimensional vector at the time $t$ within one CP-extended OFDM block

\subsection{Wideband Receiver Model}

We consider a frequency-selective wideband MU-MIMO channel between the $N$ physical transmit ports and the $U$ single-antenna users. Let $\mathbf h_{u,\tau}\in\mathbb C^{1\times N}$ denote the $\tau$-th discrete-time channel tap between the BS and user $u$, where $\tau=0,\ldots,D-1$ and $D$ is the maximum channel delay spread normalized by the sampling interval. The received signal at user $u$ before CP removal is given by the linear convolution
\begin{equation}
    y_u[t]
    =
    \sum_{\tau=0}^{D-1}
    \mathbf h_{u,\tau}\,\mathbf{x}_{\mathrm t}[t-\tau]
    +
    z_u[t],
    \label{eq:yu_time_revised}
\end{equation}
where $z_u[t]\sim\mathcal{CN}(0,\sigma^2)$ denotes the additive white Gaussian noise (AWGN) at the receiver and $\sigma^2$ is the noise variance.

Stacking the received signals of all $U$ users yields
\begin{equation}
    \mathbf{y}_{\mathrm t}[t]
    =
    \sum_{\tau=0}^{D-1}
    \mathbf H_{\tau}\mathbf x_t[t-\tau]
    +
    \mathbf{z}_{\mathrm t}[t],
    \label{eq:yt_vector_revised}
\end{equation}
where $\mathbf{H}_{\tau} \triangleq \big[ \mathbf{h}_{1,\tau}^{\top}, \mathbf{h}_{2,\tau}^{\top}, \ldots, \mathbf{h}_{U,\tau}^{\top} \big]^{\top} \in\mathbb{C}^{U\times N}$ is the aggregated channel tap matrix, and $\mathbf{z}_{\mathrm t}[t]\sim\mathcal{CN}(\mathbf{0},\sigma^2\mathbf{I}_U)$.

Assuming perfect synchronization, after standard CP removal and $K$-point DFT, the corresponding frequency-domain channel vector of user $u$ on subcarrier $k$ is obtained as
\begin{equation}
    \mathbf{h}_u[k]
    \triangleq
    \sum_{\tau=0}^{D-1}
    \mathbf{h}_{u,\tau} \,e^{-j\frac{2\pi k \tau}{K}},
    \quad
    k\in\mathcal{K}.
    \label{eq:huk_dft_def_revised}
\end{equation}

Stacking the user channels yields the wideband frequency-domain channel matrix:
\begin{equation}
    \mathbf{H}[k]
    \triangleq
    \begin{bmatrix}
        \mathbf{h}_1^{\top}[k],~
        \mathbf{h}_2^{\top}[k],~
        \ldots,~
        \mathbf{h}_U^{\top}[k]
    \end{bmatrix}^{\top}
    \in\mathbb{C}^{U\times N}.
    \label{eq:Hk_revised_final}
\end{equation}

Finally, under ideal OFDM materialization by SIM$_2$, the cascaded wave-domain modulation and receiver-side OFDM demodulation recover the standard subcarrier-decoupled representation of the frequency-selective channel. The resulting nominal frequency-domain received signal model is
\begin{align}
    \mathbf{y}[k]
    &=
    \mathbf{H}[k]\mathbf{x}[k]
    +
    \mathbf{n}[k]
    \label{eq:yk_nominal_revised} \\
    &=
    \mathbf H[k]\mathbf P[k]
    \boldsymbol{\Lambda}_{\rho}[k]\mathbf s[k]
    +
    \mathbf n[k],
    \quad
    k\in\mathcal{K},
    \label{eq:yk_nominal_expanded_revised}
\end{align}
where $\mathbf{n}[k]\sim\mathcal{CN}(\mathbf{0},\sigma^2\mathbf{I}_U)$ is the frequency-domain noise vector. This nominal model serves as the ideal benchmark, whereas the actual generalized received signal suffering from SIM$_2$-induced residual cross-subcarrier coupling will be rigorously formulated in Section III.

\begin{remark}
\textnormal{The discrete-time tap model can be equivalent to a geometric wideband channel representation. Specifically, when a geometric channel model is adopted, the frequency-domain channel of user \(u\) on subcarrier \(k\) can be written as
\begin{equation}
\mathbf h_u[k]
=
\sum_{p=1}^{P_u}
g_{u,p}[k]\boldsymbol{\alpha}_{u,p}^{H}(f_k),
\quad k\in\mathcal K,
\end{equation}
where \(P_u\) is the number of dominant propagation paths, \(\boldsymbol{\alpha}_{u,p}(f_k)\in\mathbb C^{N\times1}\) is the transmit-side array steering vector at the physical subcarrier frequency \(f_k\), and \(g_{u,p}[k]\) includes the path gain, delay-dependent phase, and pulse-shaping response.}
\end{remark}
\vspace{-0.3 cm}
\section{Problem Formulation}
Based on the system model in Section~II, the proposed architecture leads to a two-stage optimization procedure with different timescales. Specifically, the residual non-ideal behavior of SIM$_2$ induces a fixed coupling operator due to modulation, which is explicitly embedded into the end-to-end system model. Under this fixed coupling pattern, SIM$_1$ is then optimized online in a communication-oriented manner by jointly updating its layer-wise phase responses and the integrated symbol-loading power coefficients.
\vspace{-0.3 cm}
\subsection{Offline SIM$_2$ Design for Wave-Domain OFDM Modulation}

Since the CP-OFDM modulation standard is independent of the instantaneous channel state information (CSI), the design of SIM$_2$ is defined as a process of wave-domain algorithmic OFDM modulation. In this stage, the goal is to compile the mathematical multi-port CP-extended IDFT operator $\mathbf{Q}_{\mathrm{ideal}}\in\mathbb{C}^{NT\times NK}$ into the EM propagation properties of the metasurface. The offline SIM$_2$ design problem is formulated as minimizing the operator fitting error.

We first formalize the ideal OFDM modulation operator that SIM$_2$ aims to realize in the wave domain. Let $\mathbf{F}_{\mathrm{IDFT}}\in\mathbb{C}^{K\times K}$ denote the normalized IDFT matrix with entries
\begin{equation}
    [\mathbf{F}_{\mathrm{IDFT}}]_{a,b}
    =
    \frac{1}{\sqrt{K}}e^{j\frac{2\pi ab}{K}},
    \qquad
    a,b\in\mathcal{K}.
    \label{eq:FIDFT_sec3}
\end{equation}

To append a CP, define the CP-insertion matrix
\begin{equation}
    \mathbf{P}_{\mathbf{CP}}
    =
    \begin{bmatrix}
        \mathbf{0}_{N_{\mathrm{CP}}\times (K-N_{\mathrm{CP}})} & \mathbf{I}_{N_{\mathrm{CP}}} \\
        \mathbf{I}_{K}
    \end{bmatrix}
    \in\{0,1\}^{T\times K}.
    \label{eq:Ccp_sec3}
\end{equation}

\begin{figure}
	\centerline{\includegraphics[width=0.43\textwidth]{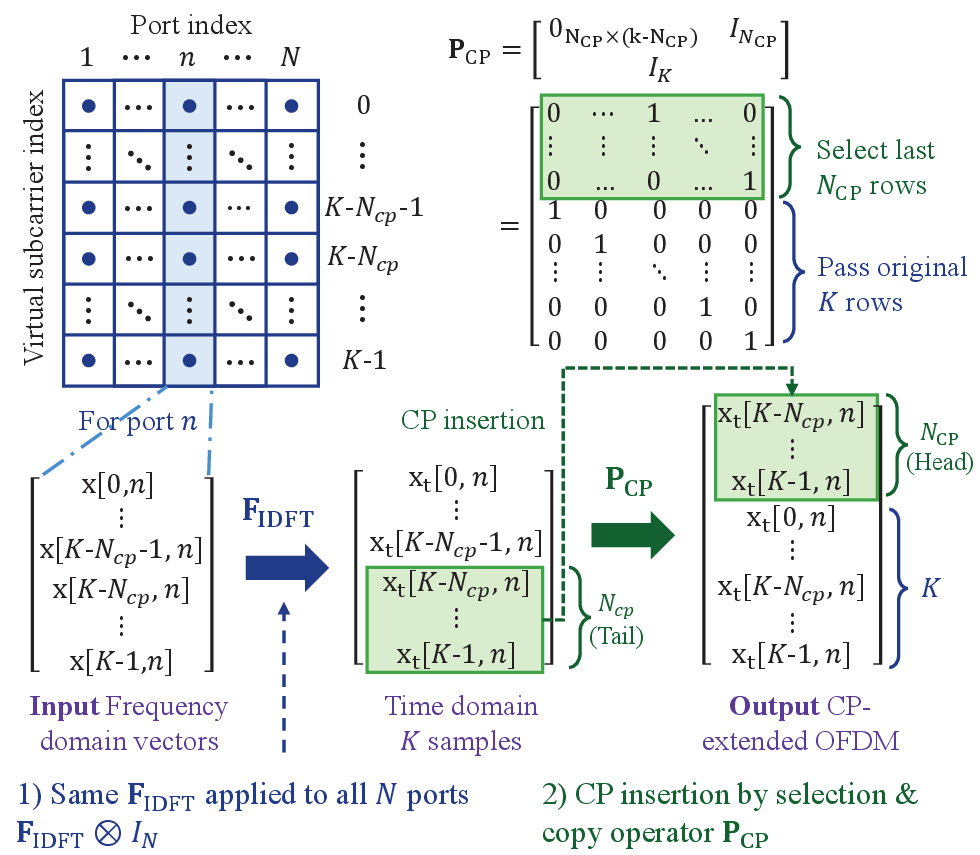}}
	\caption{Schematic illustration of the SIM$_2$-based wave-domain realization of IDFT and CP insertion.}
	\label{Fig1}
\vspace{-0.5 cm}
\end{figure}

The ideal CP-extended OFDM modulation operator for a single transmit port is therefore $\mathbf{F}_{\mathrm{CP}} \triangleq \mathbf{P}_{\mathbf{CP}}\mathbf{F}_{\mathrm{IDFT}} \in\mathbb{C}^{T\times K}$. Applying this uniform modulation across all $N$ physical transmit ports yields the ideal multiport operator:
\begin{align}
    \mathbf{Q}_{\mathrm{ideal}}
    &=
    \mathbf{F}_{\mathrm{CP}}\otimes\mathbf{I}_{N}
    \label{eq:Qideal_sec3} \\
    &=
    \begin{bmatrix}
        \mathbf{Q}_{\mathrm{ideal}}^{\top}[0],~
        \mathbf{Q}_{\mathrm{ideal}}^{\top}[1],~
        \ldots,~
        \mathbf{Q}_{\mathrm{ideal}}^{\top}[T-1]
    \end{bmatrix}^{\top},
\end{align}
where each $\mathbf{Q}_{\mathrm{ideal}}[t]\in\mathbb{C}^{N\times NK}$ denotes the desired wave-domain modulation mapping at time index $t$. Fig.~\ref{Fig1} explains why SIM2 can be interpreted as a wave-domain compiler of the conventional CP-OFDM modulation rule.

In practical implementation, SIM$_2$ can only approximate $\mathbf{Q}_{\mathrm{ideal}}$ under finite metasurface dimensions and discrete phase constraints. To account for possible row-wise complex mismatch, we introduce the auxiliary diagonal matrix
\begin{equation}
    \mathbf{D}_{t}(\boldsymbol{\beta}_{t})
    \triangleq
    \operatorname{diag}(\beta_{t,1},\beta_{t,2},\ldots,\beta_{t,N})
    \in\mathbb{C}^{N\times N},
    \label{eq:Dt_beta_sec3}
\end{equation}
where \(\boldsymbol{\beta}_t\in\mathbb C^{N\times1}\) collects the calibrated output-port scaling coefficients for time sample \(t\). These coefficients are determined offline and fixed together with SIM$_2$. They do not depend on the transmitted data symbols or instantaneous CSI, and their induced gain is included in the transmit-power normalization used for all compared architectures. Then the offline SIM$_2$ design problem is formulated as
\begin{subequations}
\begin{align}
(\mathcal{P}_1)~\min_{\bar{\mathbf{\Psi}},\,\{\boldsymbol{\beta}_{t}\}}
    \quad
    &
    \sum_{t=0}^{T-1}
    \left\|
        \mathbf{D}_{t}(\boldsymbol{\beta}_{t})
        \mathbf{Q}({\mathbf{\Psi}}[t])
        -
        \mathbf{Q}_{\mathrm{ideal}}[t]
    \right\|_{F}^{2}
    \label{prob:SIM2_obj_sec3}
    \\
    \text{s.t.}\quad
    &
    e^{j\zeta_{\mu}^{\ell}[t]}
    \in
    \mathcal{C}_{\mathrm{SIM2}},
    \quad
    \forall\,\ell,\mu,t,
    \label{prob:SIM2_const_sec3}
\end{align}
\label{prob:SIM2_sec3}
\end{subequations}
\!\!where $\mathcal{C}_{\mathrm{SIM2}}$ denotes the feasible discrete phase set of SIM$_2$. Since \eqref{prob:SIM2_sec3} depends only on the target OFDM waveform structure and the hardware constraints of SIM$_2$, it is solved offline and reused during online transmission.
\vspace{-0.3 cm}
\subsection{SIM$_2$-Induced Effective OFDM-Domain Coupling}
After obtaining the offline SIM$_2$ solution, we characterize its residual effect at the receiver after standard OFDM demodulation. Assuming the CP length satisfies $ N_{\mathrm{CP}}\ge D-1$, the inter-block interference is removed after CP stripping, and the remaining convolution becomes circular over each OFDM block. Define the CP-removal matrix
\begin{equation}
    \mathbf{R}_{\mathrm{cp}}
    =
    \begin{bmatrix}
        \mathbf{0}_{K\times N_{\mathrm{CP}}} & \mathbf{I}_{K}
    \end{bmatrix}
    \in\{0,1\}^{K\times T},
    \label{eq:Rcp_revised_final}
\end{equation}
and $\mathbf{F}_{\mathrm{DFT}}\in\mathbb{C}^{K\times K}$ is the normalized $K$-point DFT matrix. 

After the row-wise calibration, the effective SIM$_2$ operator at time sample $t$ is defined as
\begin{align}
\widetilde{\mathbf Q}(\mathbf\Psi[t],\boldsymbol\beta_t)
=
\mathbf D_t(\boldsymbol\beta_t)\mathbf Q(\mathbf\Psi[t]).
\end{align}

Stacking all time samples gives
\begin{align}
\widetilde{\mathbf Q}(\bar{\mathbf\Psi},\bar{\boldsymbol\beta})
=
[
\widetilde{\mathbf Q}^{\mathsf T}(\mathbf\Psi[0],\boldsymbol\beta_0),
\ldots,
\widetilde{\mathbf Q}^{\mathsf T}(\mathbf\Psi[T-1],\boldsymbol\beta_{T-1})
]^{\mathsf T}.
\end{align}

To explicitly characterize the residual OFDM-domain coupling induced by a nonideal SIM$_2$ after receiver-side CP removal and DFT, define the effective operator
\begin{align}
    \mathbf{T}(\bar{\mathbf\Psi},\bar{\boldsymbol\beta})
    \triangleq
(\mathbf F_{\rm DFT}\mathbf R_{\rm CP}\otimes\mathbf I_N)
\widetilde{\mathbf Q}(\bar{\mathbf\Psi},\bar{\boldsymbol\beta})
    \in\mathbb{C}^{NK\times NK}.
    \label{eq:T_operator_revised_final}
\end{align}

Substituting the optimized configurations $\bar{\mathbf{\Psi}}^{\star}$ and $\bar{\boldsymbol\beta}^\star$ from \eqref{prob:SIM2_sec3} yields the fixed coupling operator $\mathbf{T}^{\star} \triangleq \mathbf{T}(\bar{\mathbf{\Psi}}^{\star},\bar{\boldsymbol\beta}^\star)$. Writing $\mathbf{T}^{\star}\in\mathbb{C}^{NK\times NK}$ in a block-wise manner:
\begin{equation}
    \mathbf{T}^{\star}
    =
    \begin{bmatrix}
        \mathbf{T}^{\star}[0,0] & \cdots & \mathbf{T}^{\star}[0,K-1]\\
        \vdots & \ddots & \vdots\\
        \mathbf{T}^{\star}[K-1,0] & \cdots & \mathbf{T}^{\star}[K-1,K-1]
    \end{bmatrix},
    \label{eq:Tstar_sec3}
\end{equation}
where the diagonal block $\mathbf{T}^{\star}[k,k]\in\mathbb{C}^{N\times N}$ captures the in-band mapping on subcarrier $k$, and the off-diagonal blocks $\mathbf{T}^{\star}[k,i]$ with $i\neq k$ quantify the residual cross-subcarrier coupling introduced by the imperfect wave-domain OFDM materialization. Under ideal conditions, $\mathbf{T}^{\star}=\mathbf{I}_{NK}$; however, in practical SIM setups, these leakage terms must be explicitly accounted for in the online design.

Considering fitted SIM$_2$ satisfies the CP-consistency condition with small residual error. With $\mathbf{T}^{\star}$ fixed, the effective received signal model in \eqref{eq:yk_nominal_expanded_revised} can be accurately represented as
\begin{equation}
    \mathbf y[k]
    =
    \sum_{i=0}^{K-1}
    \mathbf H[k]\mathbf T^\star[k,i]\mathbf P[i]
    \boldsymbol{\Lambda}_{\rho}[i]\mathbf s[i]
    +
    \mathbf n[k],
    \quad
    k\in\mathcal{K},
    \label{eq:yk_generalized_fixedT}
\end{equation}
where this effective model captures the coupled effects of integrated power loading, SIM$_1$ precoding, SIM$_2$-induced residual OFDM-domain coupling, and the physical wideband channel.

Let $\mathbf{p}_{u}[k]\in\mathbb{C}^{N\times 1}$ denote the effective beamforming vector intended for user $u$ on subcarrier $k$, and $\mathbf{h}_{u}[k]\in\mathbb{C}^{1\times N}$ denote the $u$-th row of $\mathbf{H}[k]$. Then the received signal at user $u$ on subcarrier $k$ is
\begin{align}
    y_{u}[k]
    =
    &\underbrace{
        \sqrt{\rho_{u,k}}\,
        \mathbf{h}_{u}[k]\mathbf{T}^{\star}[k,k]\mathbf{p}_{u}[k]\,s_{u}[k]
    }_{\text{Desired signal}}
    \nonumber\\
    &+
    \underbrace{
        \sum_{\substack{j=1\\j\neq u}}^{U}
         \sqrt{\rho_{j,k}}\,\mathbf{h}_{u}[k]\mathbf{T}^{\star}[k,k]\mathbf{p}_{j}[k]\,s_{j}[k]
    }_{\text{Same-subcarrier MUI}}
    \nonumber\\
    &+
    \underbrace{
        \sum_{\substack{i=0\\i\neq k}}^{K-1}
        \sum_{j=1}^{U}
        \sqrt{\rho_{j,i}}\,
        \mathbf{h}_{u}[k]\mathbf{T}^{\star}[k,i]\mathbf{p}_{j}[i]\,s_{j}[i]
    }_{\text{Cross-subcarrier ICI}}
    +
    n_{u}[k],
    \label{eq:yu_split_sec3}
\end{align}
where $n_u[k]\sim\mathcal{CN}(0,\sigma^2)$. Consequently, the corresponding signal-to-interference-plus-noise ratio (SINR) for user $u$ on subcarrier $k$ is given by \eqref{eq:SINR_sec3}. Here, the first interference term is mainly associated with residual multiuser interference (MUI) after SIM$_1$ precoding, whereas the second term is caused by cross-subcarrier leakage induced by the nonideal SIM$_2$ coupling matrix.

\begin{figure*}[!t]
\centering
\begin{equation}
\gamma_u[k]
=
\frac{
\rho_{u,k}
\left|
\mathbf h_u[k]\mathbf T^{\star}[k,k]\mathbf p_u[k]
\right|^2
}{
\sum_{\substack{j=1\\j\ne u}}^{U}
\rho_{j,k}
\left|
\mathbf h_u[k]\mathbf T^{\star}[k,k]\mathbf p_j[k]
\right|^2
+
\sum_{\substack{i=0\\i\ne k}}^{K-1}
\sum_{j=1}^{U}
\rho_{j,i}
\left|
\mathbf h_u[k]\mathbf T^{\star}[k,i]\mathbf p_j[i]
\right|^2
+
\sigma^2
}.
    \label{eq:SINR_sec3}
\end{equation}
\hrule 
\vspace{-0.5 cm}
\end{figure*}

\subsection{Joint SIM$_1$ Phase and Power Optimization}

The phase shifts $\mathbf{\Phi}$ and integrated symbol-loading power coefficients $\boldsymbol\rho=\{\rho_{u,k}\}_{u\in\mathcal U,k\in\mathcal K}$ are jointly optimized according to the end-to-end communication objective under the actual wideband channel and the fixed SIM$_2$-induced coupling pattern. Define the achievable sum spectral efficiency
\begin{equation}
    R(\mathbf{\Phi},\boldsymbol{\rho};\mathbf{T}^{\star})
    \triangleq
    \frac{1}{K+N_{\mathrm{CP}}}\,
    \sum_{k=0}^{K-1}\sum_{u=1}^{U}
    \log_{2}\!\left(1+\gamma_{u}[k]\right).
    \label{eq:Rsum_sec3}
\end{equation}

Then the SIM$_1$ design problem is formulated as
\begin{subequations}
\begin{align}
(\mathcal P_2):\quad
\max_{\mathbf\Phi,\boldsymbol\rho}\quad&
R(\mathbf\Phi,\boldsymbol\rho;\mathbf T^\star)
\\
\mathrm{s.t.}\quad&
e^{j\theta_m^l}\in\mathcal C_{\mathrm{SIM1}},
\quad \forall l\in\mathcal L_1,\; m\in\mathcal D_l,
\\
&
\rho_{u,k}\ge 0,\quad \forall u\in\mathcal U,\; k\in\mathcal K,     \label{prob:SIM1_rho_nonneg_sec3}
\\
&
\sum_{k=0}^{K-1}\sum_{u=1}^{U}\rho_{u,k}
\le P_{\mathrm{tot}},
    \label{prob:SIM1_power_sec3}
\end{align}
\label{prob:SIM1_joint_sec3}
\end{subequations}
\!\!where $\mathcal{C}_{\mathrm{SIM1}}$ denotes the feasible discrete phase set of SIM$_1$, and $P_{\mathrm{tot}}$ is the total power budget injected at the symbol-loading interface. Problem \eqref{prob:SIM1_joint_sec3} constitutes a challenging, highly coupled non-convex optimization. However, unlike conventional preliminary operator-fitting approaches, solving \eqref{prob:SIM1_joint_sec3} explicitly adapts the physical behavior of SIM$_1$ to the true end-to-end communication environment.

\section{Proposed Optimization Framework}

In this section, a unified two-stage alternating-optimization (AO) framework is used for the proposed architecture. We first present the offline synthesis of SIM$_2$ to iteratively obtain the reusable phase shifts and the fixed OFDM-domain coupling matrix $\mathbf{T}^{\star}$. Then we formulate the online joint optimization of the integrated power-loading coefficients and the multilayer phase shifts of SIM$_1$, aiming to maximize the sum spectral efficiency under the residual coupling $\mathbf{T}^{\star}$.

\subsection{Offline Optimization of SIM$_2$}
\label{IV.A}

For fixed $\bar{\mathbf{\Psi}}$, problem \eqref{prob:SIM2_sec3} decouples across time indices and output ports. Let $[\mathbf Q(\mathbf\Psi[t])]_{n,:}$ and $[\mathbf Q_{\mathrm{ideal}}[t]]_{n,:}$ denote the $n$-th rows of the actual and ideal instantaneous modulation operators at time index $t$, respectively. By differentiating \eqref{prob:SIM2_obj_sec3} with respect to $\beta_{t,n}^{*}$, the optimal complex row-wise scaling coefficient is given by
\begin{equation}
    \beta_{t,n}^{\star}
    =
    \begin{cases}
    \dfrac{
        [\mathbf Q_{\mathrm{ideal}}[t]]_{n,:}
        [\mathbf Q(\mathbf\Psi[t])]_{n,:}^{H}
    }{
        \|[\mathbf Q(\mathbf\Psi[t])]_{n,:}\|_2^2
    },
    & \|[\mathbf Q(\mathbf\Psi[t])]_{n,:}\|_2^2 > 0,\\[2ex]
    0, & \text{otherwise}.
    \end{cases}
    \label{eq:beta_opt_sec4}
\end{equation}

To optimize the multilayer discrete phase shifts, we propose a two-stage solver comprising a continuous-phase Adam-PGD warm-up and a coordinate-wise exact discrete refinement. Let $\mathbf{v}_{\ell}[t] \triangleq [e^{j\zeta_1^{\ell}[t]},\ldots,e^{j\zeta_{\bar d_\ell}^{\ell}[t]}]^{\top}$ denote the phase vector, such that $\mathbf{\Psi}^{\ell}[t]=\operatorname{diag}(\mathbf{v}_{\ell}[t])$. 

We first relax the discrete codebook constraints to evaluate the gradients in the continuous angle domain. For the $\ell$-th layer at time index $t$, define the forward cascaded matrix $\mathbf{B}_{\ell}[t]$ and the backward cascaded matrix $\mathbf{A}_{\ell+1}[t]$ as
\begin{align}
    \mathbf{B}_{\ell}[t] &\triangleq \mathbf{U}^{\ell}\mathbf{\Psi}^{\ell-1}[t]\mathbf{U}^{\ell-1}\cdots\mathbf\Psi_1[t]\mathbf{U}^1, \\
    \mathbf{A}_{\ell+1}[t] &\triangleq \mathbf{D}_{t}(\boldsymbol{\beta}_{t})\mathbf{\Psi}^{L_2}[t]\mathbf U^{L_2}\cdots\mathbf{\Psi}^{\ell+1}[t]\mathbf{U}^{\ell+1},
\end{align}
with $\mathbf{B}_{1}[t] = \mathbf{U}^1$ and $\mathbf{A}_{L_2+1}[t] = \mathbf{D}_{t}(\boldsymbol{\beta}_{t})$. Using the current fitting residual $\mathbf{E}_{t} \triangleq \mathbf{D}_{t}(\boldsymbol{\beta}_{t})\mathbf{Q}(\mathbf{\Psi}[t]) - \mathbf{Q}_{\mathrm{ideal}}[t]$, the exact phase-angle gradient of the fitting error with respect to $\boldsymbol{\zeta}_\ell[t]$ is given by the chain rule:
\begin{equation}
    \nabla_{\boldsymbol{\zeta}^{\ell}[t]}
    =
    -2\Im\!\left\{
        \mathbf{v}_{\ell}[t]\odot \operatorname{diag}\big( \mathbf{B}_{\ell}[t]\mathbf{E}_{t}^{H}\mathbf{A}_{\ell+1}[t] \big)
    \right\},
    \label{eq:sim2_grad_sec4_new}
\end{equation}
where $\boldsymbol{\zeta}^{\ell}[t]$ is updated via an Adam-based projected gradient descent (PGD) step. The adaptive momentum acts as a diagonal pre-conditioner, guiding the phase configurations toward a favorable solution region before discrete projection.

After the continuous warm-up, the phase vectors are quantized onto $\mathcal{C}_{\mathrm{SIM2}}$. To suppress the performance loss induced by direct one-shot quantization, the discrete points are further refined by an exact element-wise block coordinate descent (E-BCD). The scaled instantaneous operator can be exactly factorized at the $\ell$-th layer as $\mathbf{A}_{\ell+1}[t] \operatorname{diag}(\mathbf{v}_{\ell}[t]) \mathbf{B}_{\ell}[t]$. By vectorizing this factorization and discarding constant terms, the subproblem for $\mathbf{v}_{\ell}[t]$ is mathematically equivalent to the quadratic maximization:
\begin{equation}
    \max_{\mathbf{v}_{\ell}[t]\in\mathcal{C}_{\mathrm{SIM2}}^{\bar d_\ell}}
    -\mathbf{v}_{\ell}^{H}[t]\mathbf{R}_{\ell}[t]\mathbf{v}_{\ell}[t]
    +
    2\Re\!\left\{
        \mathbf{s}_{\ell}^{H}[t]\mathbf{v}_{\ell}[t]
    \right\},
    \label{eq:sim2_quad_sec4_new}
\end{equation}
where $\mathbf{R}_{\ell}[t] \triangleq \mathbf{C}_{\ell}^{H}[t]\mathbf{C}_{\ell}[t]\succeq \mathbf 0$, $\mathbf{s}_{\ell}[t] \triangleq \mathbf{C}_{\ell}^{H}[t]\operatorname{vec}(\mathbf{Q}_{\mathrm{ideal}}[t])$, and $\mathbf{C}_{\ell}[t] \triangleq \mathbf{B}_{\ell}^{\top}[t]\circ \mathbf{A}_{\ell+1}[t]$.

Fixing all other entries of \(\mathbf v_\ell[t]\), the unit-modulus constraint reduces the coordinate-wise quadratic subproblem to a scalar phase-alignment problem for the \(\mu\)-th meta-atom. The equivalent scalar aggregated field is given by
\begin{equation}
    z_{\ell,\mu}[t]
    =
    [\mathbf{s}_{\ell}[t]]_\mu
    -
    [\mathbf{R}_{\ell}[t]]_{\mu,:}\mathbf{v}_{\ell}[t]
    +
    [\mathbf{R}_{\ell}[t]]_{\mu,\mu}[\mathbf{v}_{\ell}[t]]_\mu.
\end{equation}

For a finite phase codebook \(\mathcal C\), the projection operator is defined as $\Pi_{\mathcal C}(e^{j\varphi})
\triangleq
\arg\min_{c\in\mathcal C}
\left|c-e^{j\varphi}\right|^2$. Then exact coordinate-wise discrete update is analytically obtained by
\begin{equation}
    [\mathbf{v}_{\ell}[t]]_\mu
    \leftarrow
    \Pi_{\mathcal{C}_{\mathrm{SIM2}}}
    \!\left(
        e^{j\angle(z_{\ell,\mu}[t])}
    \right),
    \qquad
    \mu \in \bar{\mathcal{D}}_{\ell}.
    \label{eq:sim2_ebcd_update_sec4_new}
\end{equation}

By alternately updating $\{\boldsymbol{\beta}_t\}_{t \in\mathcal{T}}$ and $\{\mathbf{v}_{\ell}[t]\}_{t \in\mathcal{T} ,\ell\in\mathcal{L}_2}$, we obtain the offline SIM$_2$ solution $\bar{\mathbf{\Psi}}^{\star}$ and the corresponding effective OFDM-domain coupling matrix $\mathbf{T}^{\star}$.

\subsection{Online Optimization of SIM$_1$ Under Fixed $\mathbf{T}^{\star}$}

With $\mathbf{T}^{\star}$ fixed, the SIM$_1$ design is obtained by maximizing the sum spectral efficiency with respect to $\mathbf{\Phi}$ and $\boldsymbol{\rho}$.

\subsubsection{FP-Based Equivalent Reformulation}

For compactness, define desired-signal amplitude on subcarrier $k$ as $ d_{u,k}(\mathbf\Phi,\boldsymbol\rho)   \triangleq \sqrt{\rho_{u,k}}\,\mathbf h_u[k]\mathbf T^\star[k,k]\mathbf p_u[k]$, and the total received signal-plus-noise power as $\Xi_{u,k}(\mathbf{\Phi},\boldsymbol{\rho}) \triangleq     \sum_{i=0}^{K-1}    \sum_{j=1}^{U}    \rho_{j,i} | \mathbf{h}_u[k]\mathbf{T}^{\star}[k,i]\mathbf{p}_j[i] |^2 + \sigma^2$. The SINR is thereby expressed as $\gamma_u[k] = |d_{u,k}|^2 / (\Xi_{u,k}-|d_{u,k}|^2)$.

To decouple the fractional objective, we apply the Lagrangian dual transform \cite{FP}. Noting the ratio structure $\gamma_u[k]/(1+\gamma_u[k]) = |d_{u,k}|^2 / \Xi_{u,k}$, we obtain
\begin{align}
    & \log_{2}\!\left(1+\gamma_u[k]\right) = \frac{1}{\ln 2} \, \ln\!\left(1+\gamma_u[k]\right) \nonumber \\
    &= \frac{1}{\ln 2}
    \max_{\eta_{u,k}\ge 0}
    \left[
        \ln(1+\eta_{u,k})-\eta_{u,k}
        +
        (1+\eta_{u,k})
        \frac{|d_{u,k}|^2}{\Xi_{u,k}}
    \right],
    \label{eq:dual_transform_sec4}
\end{align}
where $\boldsymbol{\eta}\triangleq\{\eta_{u,k}\}_{u \in\mathcal{U} ,k\in\mathcal{K}}$ are nonnegative auxiliary variables. The constant factor \(1/\ln 2\) is omitted in \(f_{\rm FP}\) since it does not affect the optimizer. For fixed $\mathbf{\Phi}$ and $\boldsymbol{\rho}$, the optimal $\eta_{u,k}$ is explicitly given by
\begin{equation}
    \eta_{u,k}^{\star}
    =
    \gamma_u[k].
    \label{eq:eta_opt_sec4}
\end{equation}

The remaining fractional term in \eqref{eq:dual_transform_sec4} is further decoupled via the multidimensional complex quadratic transform
\begin{align}
    &(1+\eta_{u,k})\frac{|d_{u,k}|^2}{\Xi_{u,k}} \nonumber \\
    &=
    \max_{y_{u,k}\in\mathbb{C}}
    \left[
        2\sqrt{1+\eta_{u,k}}\,\Re\{y^{*}_{u,k}d_{u,k}\}
        -
        |y_{u,k}|^2\Xi_{u,k}
    \right].
    \label{eq:qt_identity_sec4}
\end{align}

By introducing the complex auxiliary variables $\mathbf{y}\triangleq\{y_{u,k}\}_{u \in\mathcal{U} ,k\in\mathcal{K}}$, the highly coupled original problem is rigorously converted into the equivalent objective:
\begin{equation}
(\mathcal{P}_6)~\max_{\mathbf{\Phi},\,\boldsymbol{\rho},\,\boldsymbol{\eta},\,\mathbf{y}}
    \quad
    f_{\mathrm{FP}}(\mathbf{\Phi},\boldsymbol{\rho},\boldsymbol{\eta},\mathbf{y}),
    \label{eq:FP_master_sec4}
\end{equation}
where
\begin{align}
    f_{\mathrm{FP}}
    =
    \sum_{k=0}^{K-1}\sum_{u=1}^{U}
    \Big[
        &\ln(1+\eta_{u,k})-\eta_{u,k}
        -
        |y_{u,k}|^2\Xi_{u,k}
        \nonumber\\
        &+
        2\sqrt{1+\eta_{u,k}}\,
        \Re\!\left\{
            y_{u,k}^{*}d_{u,k}
        \right\}
    \Big].
    \label{eq:FP_obj_sec4}
\end{align}

For fixed $\mathbf{\Phi}$, $\boldsymbol{\rho}$, and $\boldsymbol{\eta}$, the optimal quadratic-transform variable admits the closed-form expression
\begin{equation}
    y_{u,k}^{\star}
    =
    \frac{
        \sqrt{1+\eta_{u,k}}\,
        d_{u,k}
    }{
        \Xi_{u,k}
    }.
    \label{eq:y_opt_sec4}
\end{equation}

Therefore, the original problem $\mathcal{P}_2$ is equivalent to solving problem $\mathcal{P}_6$ by alternately updating the auxiliary blocks $(\boldsymbol{\eta},\mathbf{y})$, the power coefficients $\boldsymbol{\rho}$, and the phase shifts $\mathbf{\Phi}$.

\subsubsection{Power-Loading Update}

Fixing $\mathbf{\Phi}$, $\boldsymbol{\eta}$, and $\mathbf{y}$, we isolate the terms dependent on $\boldsymbol{\rho}$ in \eqref{eq:FP_obj_sec4}. Substituting the definitions of $d_{u,k}$ and $\Xi_{u,k}$, the power-allocation subproblem becomes
\begin{align}
(\mathcal{P}_7)~\max_{\boldsymbol{\rho}}
    \quad
\sum_{k=0}^{K-1}\sum_{u=1}^{U}
\left(
a_{u,k}\sqrt{\rho_{u,k}}
-
b_{u,k}\rho_{u,k}
\right),
\label{prob:power_sec4_rho}
\end{align}
where the linear and quadratic coefficients are derived as
\begin{align}
a_{u,k}
=
2\sqrt{1+\eta_{u,k}}\,
\Re\left\{
y_{u,k}^{*}
\mathbf h_u[k]\mathbf T^\star[k,k]\mathbf p_u[k]
\right\},\\
b_{u,k}
=
\sum_{r=0}^{K-1}\sum_{q=1}^{U}
|y_{q,r}|^2
\left|
\mathbf h_q[r]\mathbf T^\star[r,k]\mathbf p_u[k]
\right|^2.
\end{align}

Using the KKT conditions \cite{KKT}, we introduce the Lagrangian
    \begin{align}
    \mathcal L    =
    &
    \sum_{k=0}^{K-1}
    \sum_{u=1}^{U}
    \left(
        a_{u,k}\sqrt{\rho_{u,k}}
        -
        b_{u,k}\rho_{u,k}
    \right) \nonumber    \\ 
    &
    -
    \lambda
    \left(
        \sum_{k=0}^{K-1}\sum_{u=1}^{U}
        \rho_{u,k}
        -
        P_{\mathrm{tot}}
    \right)
    +
    \sum_{k=0}^{K-1}
    \sum_{u=1}^{U}
    \mu_{u,k}\rho_{u,k},
    \label{eq:power_lagrangian_stream}
    \end{align}
where $\lambda\ge 0$ is the Lagrange multiplier associated with the total
power constraint, and $\mu_{u,k}\ge 0$ corresponds to the nonnegativity
constraint of $\rho_{u,k}$.

For an active stream-subcarrier pair with $\rho_{u,k}>0$, the
complementary-slackness condition gives $\mu_{u,k}\!=\!0$, so that ${\partial\mathcal L}/{\partial \rho_{u,k}}    =    0$ yields $
    {\rho_{u,k}^{\star}(\lambda)}
    =
    (\frac{a_{u,k}}{2(b_{u,k}+\lambda)})^2$. If $a_{u,k}\le 0$, allocating power to the corresponding stream-subcarrier pair cannot increase the transformed objective, and hence the optimal solution is $\rho_{u,k}^{\star}=0$. Taking the nonnegativity constraint into account, the optimal power-loading coefficient is
\begin{equation}
    \rho_{u,k}^{\star}(\lambda)
    =
    \left(
        \max
        \left\{
            0,\,
            \frac{a_{u,k}}{2(b_{u,k}+\lambda)}
        \right\}
    \right)^2,
    \label{eq:rho_uk_opt}
\end{equation}
where the global water-level parameter $\lambda$ can be found via a bisection search to satisfy $\sum_{k=0}^{K-1}\sum_{u=1}^{U}\rho_{u,k}^{\star}(\lambda) \le P_{\mathrm{tot}}$.

\subsubsection{Layer-Wise Phase Optimization of SIM$_1$}

We adopt a two-step layer-wise phase update, following the same continuous-relaxation and discrete-refinement philosophy used for SIM$_2$. For the $l$-th layer, define the phase vector $\mathbf{v}_{l} \triangleq [e^{j\theta_1^{l}},\ldots,e^{j\theta_{d_l}^{l}}]^{\top}$, such that $\mathbf{\Phi}^{l}=\operatorname{diag}(\mathbf{v}_{l})$. Fixing all other phase matrices, the overall SIM$_1$ operator can be factorized as
\begin{equation}
    \mathbf{P}(\mathbf{\Phi})
    =
    \mathbf{M}_{l}\,
    \operatorname{diag}(\mathbf{v}_{l})\,
    \mathbf{N}_{l},
    \label{eq:sim1_factor_sec4}
\end{equation}
where two terms $\mathbf{M}_{l} = \mathbf{\Phi}^{L_1}\mathbf{W}^{L_1}\cdots\mathbf{\Phi}^{l+1}\mathbf{W}^{l+1}$ and $\mathbf{N}_{l} = \mathbf{W}^{l}\mathbf{\Phi}^{l-1}\mathbf{W}^{l-1}\cdots\mathbf{\Phi}^{1}\mathbf{W}^{1}$ denote the cascaded operators after and before the \(l\)-th phase layer, respectively. Empty products are interpreted as identity matrices.

To extract the effective precoding vector for user $j$ on subcarrier $i$, we define the spatial block selection matrix $\mathbf{S}_i \in \{0,1\}^{N\times NK}$ and the canonical input selector $\mathbf{e}_{i,j}\in\{0,1\}^{SK\times 1}$. By setting $\mathbf{d}_{i,j}^{(l)} \triangleq \mathbf{N}_l \mathbf{e}_{i,j}$, the fully-coupled equivalent channel is linearly mapped to $\mathbf{v}_l$:
\begin{equation}
    \mathbf{p}_{j}[i]
    =
    \mathbf{S}_{i}\mathbf{M}_{l}
    \operatorname{diag}\!\big(\mathbf{d}_{i,j}^{(l)}\big)
    \mathbf{v}_{l}.
    \label{eq:sim1_pj_linear_sec4}
\end{equation}
Consequently, the effective cascaded channel term becomes linear with respect to $\mathbf{v}_{l}$:
\begin{equation}
    \mathbf{h}_{u}[k]\mathbf{T}^{\star}[k,i]\mathbf{p}_{j}[i]
    =
    \big(\mathbf{c}_{u,k,i,j}^{(l)}\big)^{H}\mathbf{v}_{l},
    \label{eq:sim1_cukij_sec4}
\end{equation}
where the equivalent channel aggregation vector is defined as
\begin{equation}
    \big(\mathbf{c}_{u,k,i,j}^{(l)}\big)^{H}
    \triangleq
    \mathbf{h}_{u}[k]\mathbf{T}^{\star}[k,i]\mathbf{S}_{i}\mathbf{M}_{l}
    \operatorname{diag}\!\big(\mathbf{d}_{i,j}^{(l)}\big).
    \label{eq:sim1_cukij_def_sec4}
\end{equation}

Substituting \eqref{eq:sim1_cukij_sec4} into the transformed FP objective, the highly coupled phase optimization at the $l$-th layer rigorously reduces to the following discrete quadratic subproblem:
\begin{equation}
(\mathcal{P}_8)\quad
\max_{\mathbf{v}_{l}\in\mathcal{C}_{\mathrm{SIM1}}^{d_l}}
    -\mathbf{v}_{l}^{H}\mathbf{\Gamma}_{l}\mathbf{v}_{l}
    +
    2\Re\!\left\{
        \boldsymbol{\gamma}_{l}^{H}\mathbf{v}_{l}
    \right\},
    \label{eq:sim1_quad_sec4}
\end{equation}
where the positive semi-definite matrix $\mathbf{\Gamma}_{l}$ and the linear projection vector $\boldsymbol{\gamma}_{l}$ are respectively assembled as
\begin{align}
    \boldsymbol\Gamma_l
   & =
    \sum_{k=0}^{K-1}\sum_{u=1}^{U}
    |y_{u,k}|^2
    \sum_{i=0}^{K-1}\sum_{j=1}^{U}
    \rho_{j,i}
    \mathbf c_{u,k,i,j}^{(l)}
    \left(\mathbf c_{u,k,i,j}^{(l)}\right)^H,
    \label{eq:sim1_Gamma_l_sec4} \\
    \boldsymbol\gamma_l
   & =
    \sum_{k=0}^{K-1}\sum_{u=1}^{U}
    \sqrt{(1+\eta_{u,k})\rho_{u,k}}\,
    y_{u,k}
    \mathbf c_{u,k,k,u}^{(l)} .
    \label{eq:sim1_gamma_l_sec4}
\end{align}

Relaxing $\mathbf v_l\in\mathcal C_{\mathrm{SIM1}}$ from discrete to the unit-modulus constraint $|[\mathbf v_l]_m|=1$, the phase-angle gradient of \eqref{eq:sim1_quad_sec4} is
\begin{align}
\nabla_{\boldsymbol\theta_l}
=
2\Im\left\{
\mathbf v_l^{*}\odot
\left(
\boldsymbol\gamma_l-\boldsymbol\Gamma_l\mathbf v_l
\right)
\right\},
\label{73}
\end{align}
where the continuous phase vector is updated using an Adam-based projected gradient step before discrete quantization.

Similarly, we perform an E-BCD update for non-convex constraints. By isolating the $m$-th meta-atom of the $l$-th layer and fixing all the remaining entries of $\mathbf{v}_{l}$, the unit-modulus constraint collapses the quadratic interaction term. The effective scalar coefficient for the $m$-th coordinate evaluates to
\begin{equation}
    z_{l,m}
    =
    [\boldsymbol{\gamma}_{l}]_m
    -
    [\mathbf{\Gamma}_{l}]_{m,:}\mathbf{v}_{l}
    +
    [\mathbf{\Gamma}_{l}]_{m,m}[\mathbf{v}_{l}]_m.
    \label{eq:sim1_zlm_sec4}
\end{equation}
The optimal discrete update that exactly maximizes the localized subproblem is directly obtained by mapping the continuous angle onto the nearest feasible codeword:
\begin{equation}
    [\mathbf{v}_{l}]_m
    \leftarrow
    \Pi_{\mathcal{C}_{\mathrm{SIM1}}}
    \!\left(
        e^{j\angle(z_{l,m})}
    \right),
    \qquad
    m\in\mathcal{D}_{l},
    \label{eq:sim1_ebcd_update_sec4}
\end{equation}
where sweeping all coordinates $m \in \mathcal{D}_{l}$ yields an effective discrete phase refinement for the $l$-th layer.

The detailed steps of the proposed two-stage optimization are summarized in Algorithm \ref{alg:overall}. By alternately updating the layer-wise phase vectors together with the auxiliary variables $(\boldsymbol{\eta},\mathbf{y},\boldsymbol{\rho})$, we iteratively obtain the online SIM$_1$ precoding solution under the fixed residual coupling matrix $\mathbf{T}^{\star}$.

\begin{algorithm}[t]
\caption{Proposed Two-Stage AO Framework}
\label{alg:overall}
\begin{algorithmic}[1]
\STATE \textbf{Stage 1: Offline SIM$_2$ Synthesis} \\
\STATE Initialize the phase shifts $\bar{\mathbf{\Psi}}$ and scaling factors $\{\boldsymbol{\beta}_t\}$. \\
\REPEAT
    \STATE Update scaling coefficients $\{\beta_{t,n}\}$ via \eqref{eq:beta_opt_sec4}.\\
    \FOR{$t \in \mathcal{T}$ and $\ell \in \mathcal{L}_2$}
	\STATE  Perform Adam-PGD warm-up using \eqref{eq:sim2_grad_sec4_new}.
	\STATE  Project the continuous phase vectors onto $\mathcal C_{\mathrm{SIM2}}$.
	\STATE  Update discrete phase shifts $\mathbf{v}_\ell[t]$ via \eqref{eq:sim2_ebcd_update_sec4_new}.
    \ENDFOR
\UNTIL{convergence}
\STATE Compute the fixed coupling matrix $\mathbf{T}^\star$ via \eqref{eq:T_operator_revised_final}.\\
\vspace{0.15cm}
\STATE \textbf{Stage 2: Online SIM$_1$ Optimization} \\
\STATE Given $\mathbf{T}^\star$ and current wideband channel realization $\{\mathbf{H}[k]\}_{k=0}^{K-1}$. \\
\STATE Initialize the phase shifts $\mathbf{\Phi}$ and power coefficients $\boldsymbol{\rho}$. \\
\REPEAT
    \STATE Update auxiliary variables $(\boldsymbol{\eta},\mathbf{y})$ via \eqref{eq:eta_opt_sec4} and \eqref{eq:y_opt_sec4}.\\
    \STATE Find the optimal Lagrange multiplier $\lambda$ via bisection and update the power coefficients $\boldsymbol{\rho}$ via \eqref{eq:rho_uk_opt}.\\
    \FOR{$l \in \mathcal{L}_1$}
	\STATE Perform Adam-PGD warm-up using \eqref{73}.
	\STATE Project the continuous phase vectors onto $\mathcal C_{\mathrm{SIM1}}$.
	\STATE Update discrete phase shifts $\mathbf{v}_l$ via \eqref{eq:sim1_ebcd_update_sec4}.
	\ENDFOR
\UNTIL{convergence}
\STATE \textbf{Output:} Optimized configurations $\mathbf{\Phi}^\star$, $\boldsymbol{\rho}^\star$, $\bar{\mathbf{\Psi}}^\star$, and $\bar{\boldsymbol\beta}^\star$.
\end{algorithmic}
\end{algorithm}

\vspace{-0.35 cm}
\section{Results and Analysis}

\subsection{Simulation Setup}

In this section, numerical simulations are conducted to validate the proposed fully wave-domain wideband MU-MIMO OFDM transmitter.  We consider a downlink wideband MU-MIMO OFDM system operating at $f_c=28$ GHz with bandwidth $B=100$ MHz. The BS serves $U=4$ single-antenna users with one stream per user. The number of transmit ports is $N=6$, the number of subcarriers is $K=16$, and the CP length is $N_{\mathrm{CP}}=4$. The default total transmit power is set to $P_{tot}=20$ dBm. The thermal noise power spectral density is set to $-174$ dBm/Hz, and the receiver noise figure is $7$ dB. The BS-side antenna gain is set to $G_{\mathrm{BS}}=5$ dBi, following the setting in \cite{confi}, while the UE antenna gain is set to $G_{\mathrm{UE}}=0$ dBi. The users are uniformly located between $60$ m and $120$ m. The large-scale fading follows the free-space path-loss model. The small-scale frequency-selective channel contains $D=4$ taps, with each tap generated as an independent Rayleigh fading vector normalized by $1/D$. The default SIM$_1$ configuration is selected as $L_1=7$ with hidden dimension $M_1=160$, while the default SIM$_2$ configuration is $L_2=9$ with hidden dimension $M_2=160$. The phase resolutions of SIM$_1$ and SIM$_2$ are set to $B_1=B_2=6$ bits, respectively, unless otherwise specified. 
\vspace{-0.25 cm}
\subsection{Convergence and Phase-Resolution Analysis of SIMs}
\begin{figure}[t]
    \centering
    \subfigure[\label{fig:subfig1a}]{\includegraphics[width=0.27\textwidth]{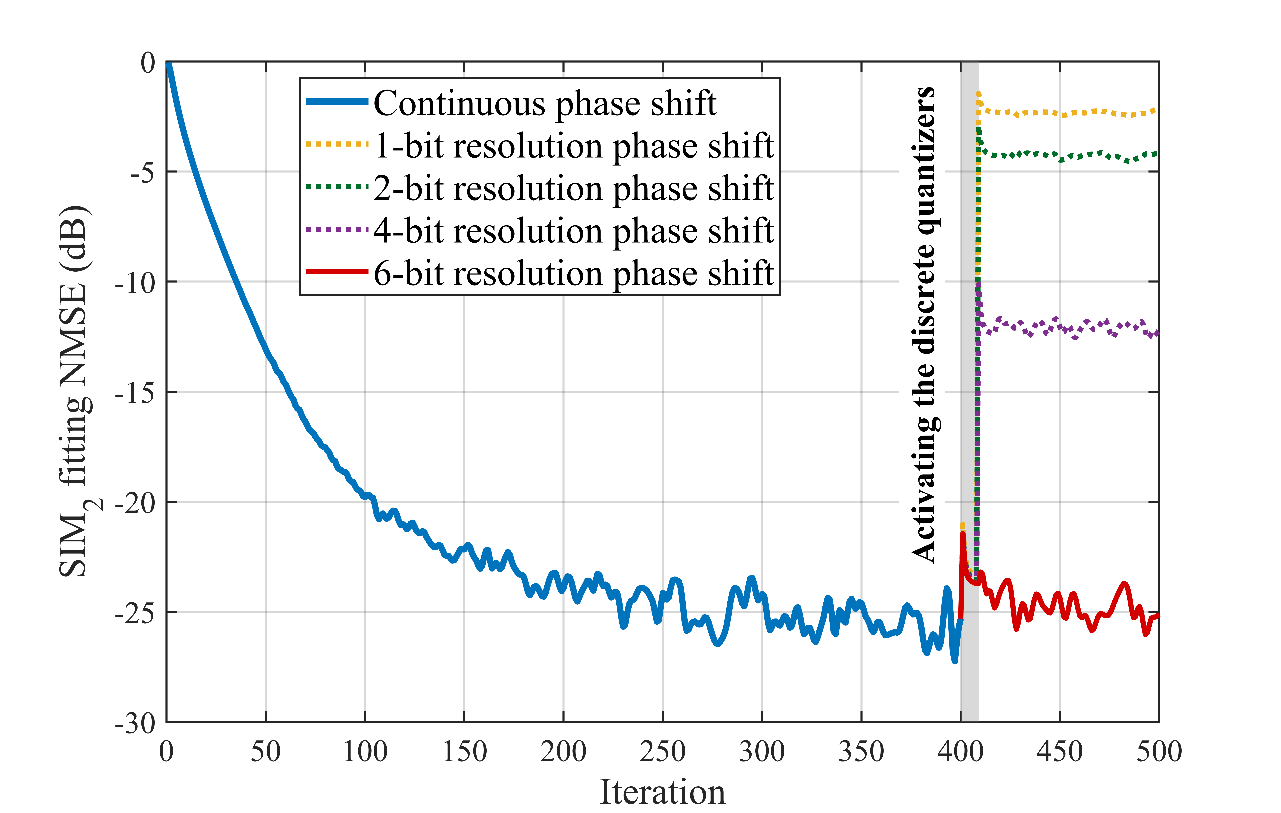}}
    \hspace{-0.6 cm}
    \subfigure[\label{fig:subfig1b}]{\includegraphics[width=0.235\textwidth]{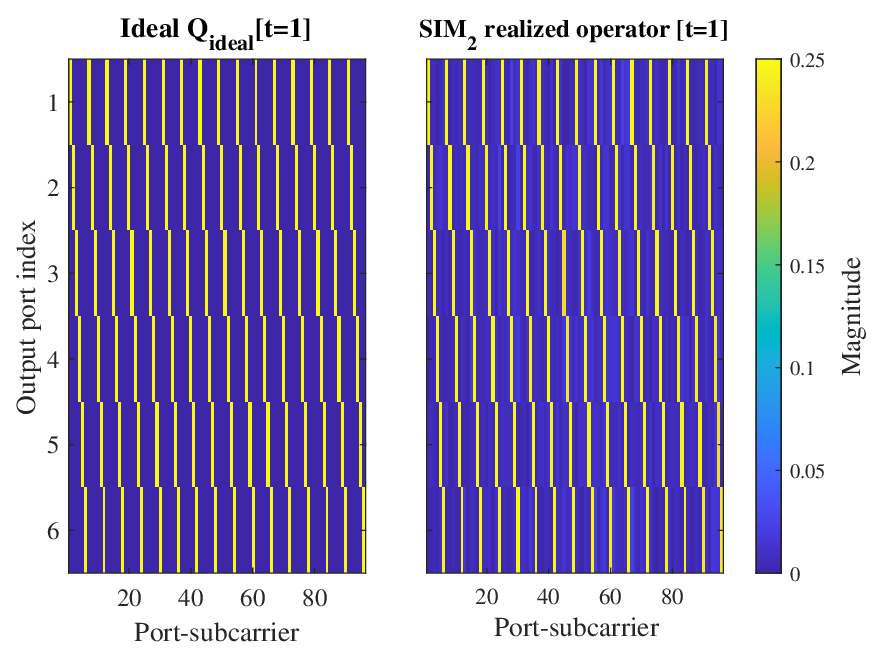}}
    \vspace{-0.2 cm}
\caption{{Validation of the offline SIM$_2$ OFDM materialization solver: (a) SIM$_2$ convergence behavior under different phase resolutions. (b) Comparison between ideal CP-OFDM and SIM$_2$.}}
    \label{Conv2}
\vspace{-0.5 cm}
\end{figure}

We first examine the convergence behavior of the proposed algorithms and the impact of finite phase resolution. Fig.~\ref{fig:subfig1a} shows the convergence of the offline SIM$_2$ OFDM materialization solver. During the continuous-phase warm-up stage, the fitting normalized mean square error (NMSE) decreases rapidly and then gradually approaches a low-error region. As the discrete quantizers are activated, different phase resolutions lead to distinct fitting floors. Low-resolution phase shifters, such as 1-bit and 2-bit, introduce severe quantization loss and therefore cannot accurately reproduce the ideal CP-OFDM operator. Increasing the resolution to 4 bits significantly reduces the fitting error, while the 6-bit curve remains very close to the continuous-phase reference even after quantization. This indicates that, under the considered configuration, 6-bit phase control provides a favorable balance between OFDM materialization accuracy and phase-control resolution. The heatmap comparisons in Fig. \ref{fig:subfig1b} further confirm that the realized operator accurately captures the sparse comb-like structure of the target IDFT-and-CP modulation operator at the representative time slot. The staggered periodic bright spots reflect the desired per-port subcarrier selectivity with only minor residual leakage.

\begin{figure}
	\centerline{\includegraphics[width=0.4\textwidth]{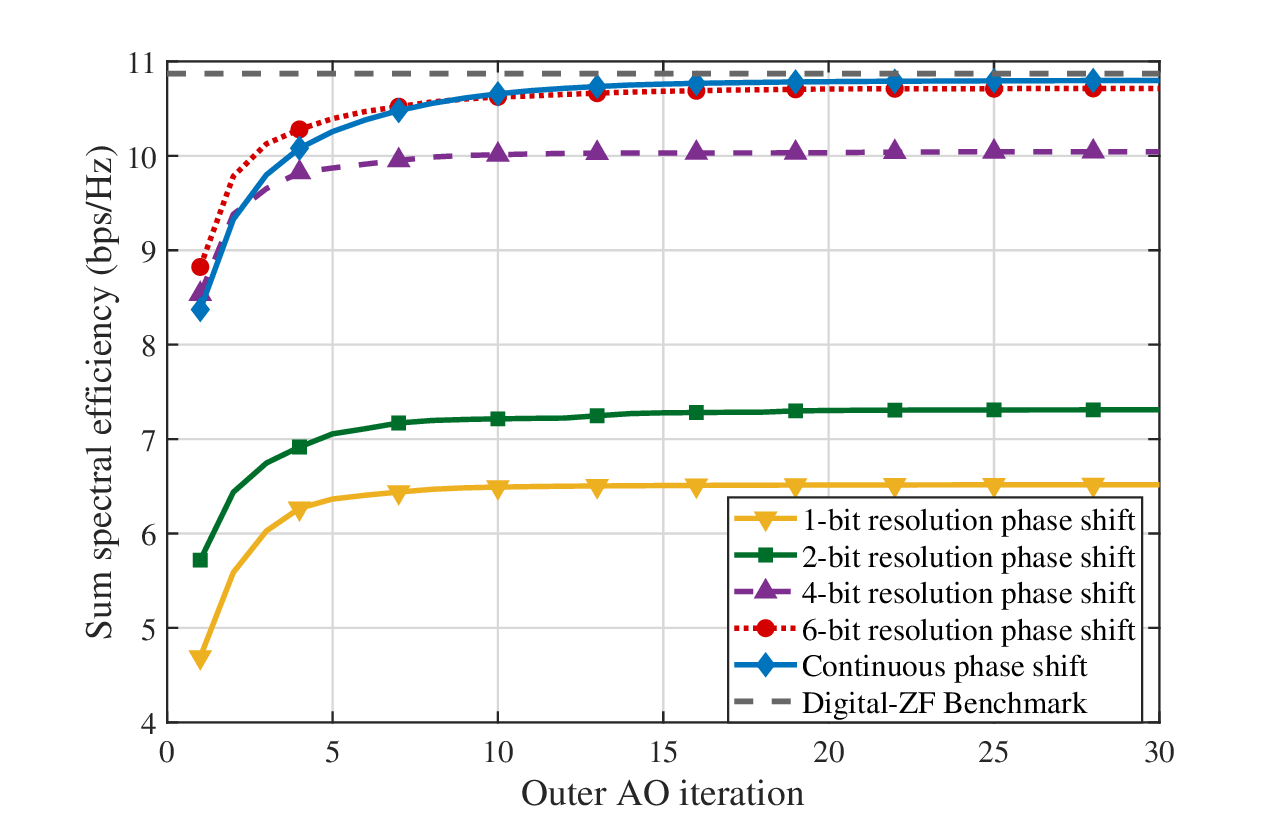}}
\caption{Convergence behavior of the online SIM$_1$ wave-domain precoding solver under different phase resolutions.}
	\label{Conv1}
\vspace{-0.6 cm}
\end{figure}

Fig.~\ref{Conv1} evaluates the online SIM$_1$ wave-domain precoding solver under ideal SIM$_2$ with $\mathbf T^\star\!=\!\mathbf I_{NK}$. The dashed curve denotes the ideal digital-ZF benchmark. All SIM$_1$ curves converge within a small number of outer AO iterations, which validates the stability of the FP/KKT power update and the layer-wise phase optimization. As the phase resolution increases, the achievable sum spectral efficiency improves monotonically. In particular, the 1-bit and 2-bit cases are limited by severe phase quantization errors, whereas the 4-bit case already achieves a substantial portion of the continuous-phase performance. The 6-bit SIM$_1$ curve nearly overlaps with the continuous-phase reference and approaches the digital-ZF benchmark. These results indicate that the proposed online solver can optimize the wave-domain MU-MIMO precoder under practical discrete phase constraints. Overall, the 6-bit setting achieves a favorable balance between implementation feasibility and algorithmic performance and is therefore adopted in the following configuration trade-off and simulations.
\vspace{-0.35 cm}
\subsection{Architecture Configuration of SIMs}
We next investigate the architecture configuration of the two SIM modules to identify a practical complexity-performance trade-off before the system evaluation. Since SIM$_1$ and SIM$_2$ implement different wave-domain functions, their configuration metrics are chosen according to their respective roles. For SIM$_2$, we evaluate the normalized OFDM materialization accuracy $\mathcal A_{\rm OFDM}=1/(1+\epsilon_{\rm SIM2})$, where $\epsilon_{\rm SIM2}=\|\mathbf T^\star-\mathbf I_{NK}\|_F^2/ \|\mathbf I_{NK}\|_F^2$, and the ICI-to-desired power ratio, which directly characterize its ability to materialize the CP-OFDM operator in the wave domain.  For SIM$_1$, we set $\mathbf T^\star=\mathbf I_{NK}$ to isolate the spatial MU-MIMO precoding capability and evaluate the normalized spectral-efficiency ratio $R_{\mathrm{SIM1}}/R_{\mathrm{DigitalZF}}$ together with the residual MUI-to-desired power ratio. 

\begin{figure}[t]
    \centering
    \subfigure[\label{SIM2a}]{\includegraphics[width=0.24\textwidth]{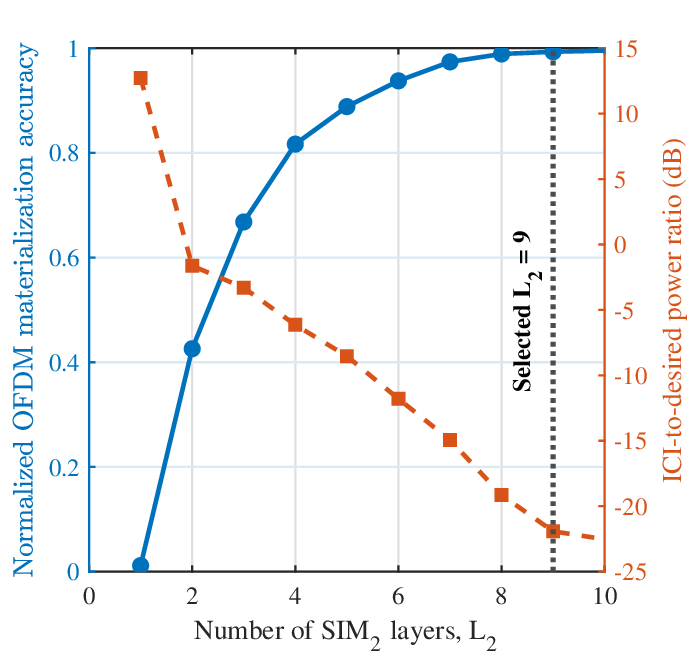}}
\hspace{-0.2 cm}
    \subfigure[ \label{SIM2b}]{\includegraphics[width=0.24\textwidth]{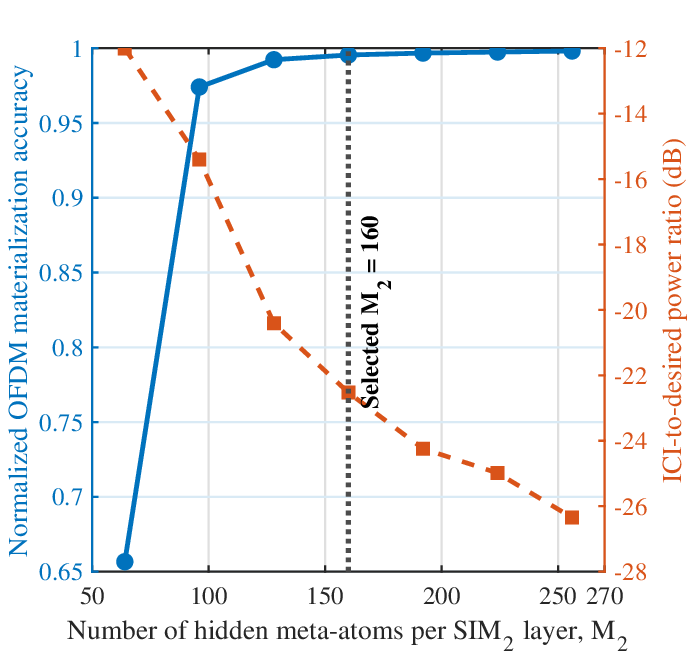}}
    \vspace{-0.2 cm}
\caption{Architecture trade-off of SIM$_2$ for wave-domain CP-OFDM modulation: (a) Impact of the number of SIM$_2$ layers $L_2$ with fixed hidden width $M_{2}\!=\!160$. (b) Impact of the hidden meta-atoms per layer with fixed depth $L_2\!=\!9$.}
    \label{SIM2}
\vspace{-0.6 cm}
\end{figure}

Fig.~\ref{SIM2} evaluates the architecture trade-off of SIM$_2$ for wave-domain CP-OFDM modulation. SIM$_2$ is responsible for the offline materialization of the IDFT-and-CP operator, and its residual error appears as cross-subcarrier ICI after receiver-side CP removal and DFT. As shown in Fig.~\ref{SIM2a}, increasing $L_2$ significantly improves the normalized accuracy and reduces the ICI-to-desired ratio. The improvement becomes much slower after $L_2\!=\!9$, indicating that the CP-OFDM modulation operator has been sufficiently approximated with the 6-bit phase resolution. Fig.~\ref{SIM2b} further shows that increasing the number of hidden meta-atoms per SIM$_2$ layer improves the OFDM fitting accuracy and reduces the residual ICI level. The gain gradually saturates for large meta-atoms around $M_{2}\!=\!160$, confirming that $M_{x,2}\!=\!10$ provides a favorable balance between fitting accuracy and hardware complexity.

\begin{figure}[t]
    \centering
    \subfigure[\label{SIM1a}]{\includegraphics[width=0.24\textwidth]{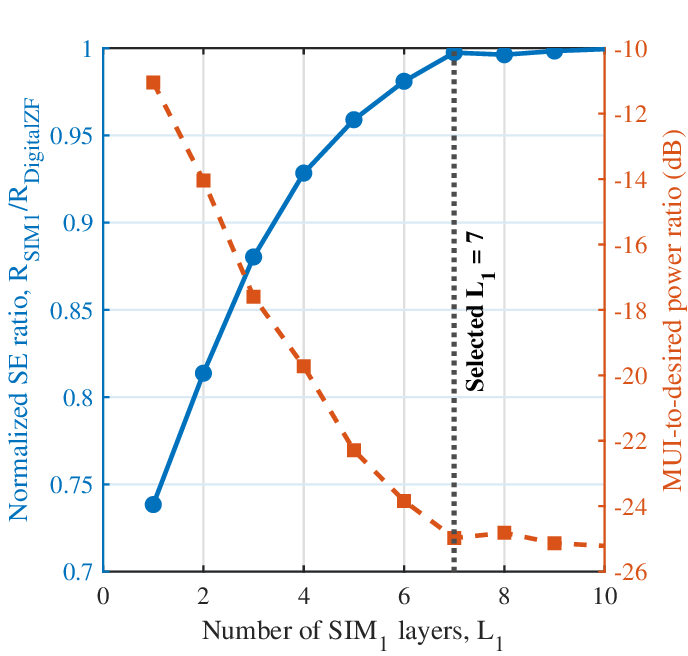}}
\hspace{-0.2 cm}
    \subfigure[ \label{SIM1b}]{\includegraphics[width=0.24\textwidth]{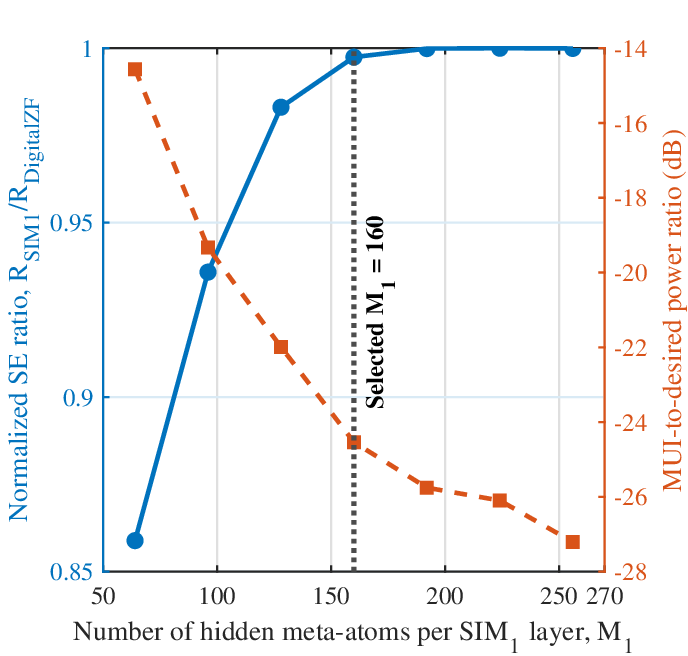}}
    \vspace{-0.2 cm}
\caption{Architecture trade-off of SIM$_1$ for wave-domain precoding under ideal SIM$_2$: (a) Impact of the number of SIM$_1$ layers $L_1$ with fixed hidden dimension $M_{1}=160$. (b) Impact of the hidden meta-atoms per layer with fixed depth $L_1=7$.}
    \label{SIM1}
\vspace{-0.5 cm}
\end{figure}

Fig.~\ref{SIM1} shows the architecture trade-off of SIM$_1$. As the number of SIM$_1$ layers increases in Fig.~\ref{SIM1a}, the normalized SE ratio improves rapidly, while the MUI-to-desired ratio decreases substantially. This is because additional layers provide more cascaded diffraction DoF, enabling stronger wave-domain spatial separation among the user streams. However, the performance gain becomes marginal after $L_1=7$, where the normalized SE ratio is already close to the digital-ZF benchmark and the residual MUI has been significantly suppressed. A similar trend is observed in Fig.~\ref{SIM1b}, when the number of layers is fixed as $L_1=7$. Increasing the hidden-layer width improves the spatial computation capability of SIM$_1$, but the gain saturates when the number of hidden meta-atoms per layer reaches $M_1=160$ with $M_{x,1}=10$ for fixed $K=16$. Therefore, we select \(L_1=7\) and \(M_1=160\) as a cost-effective SIM$_1$ configuration and use the selected layer depths and hidden dimensions in the subsequent simulations.
\vspace{-0.35 cm}
\subsection{Functional Validation of Wave-Domain CP-OFDM and Residual Coupling}
\begin{figure}[t]
    \centering
    \subfigure[\label{SIM2Oa}]{\includegraphics[width=0.295\textwidth]{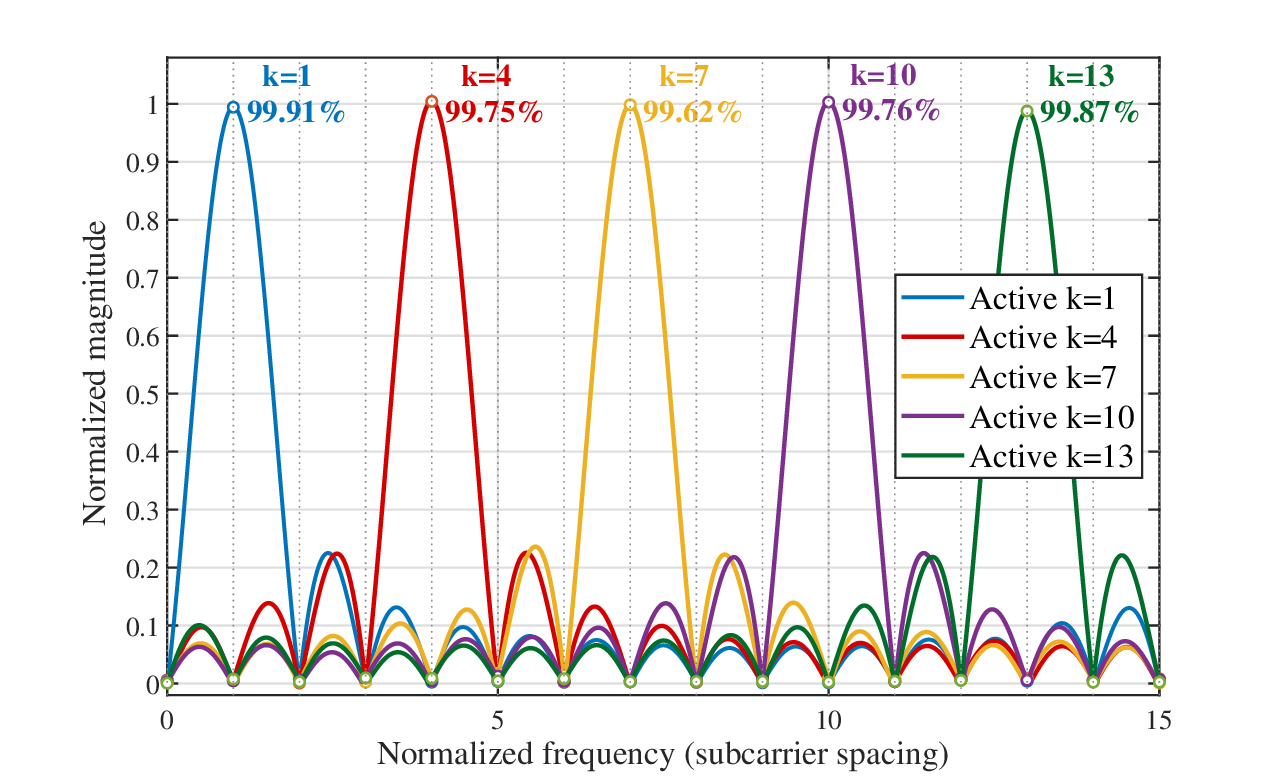}}
\hspace{-0.67 cm}
    \subfigure[ \label{SIM2Ob}]{\includegraphics[width=0.215\textwidth]{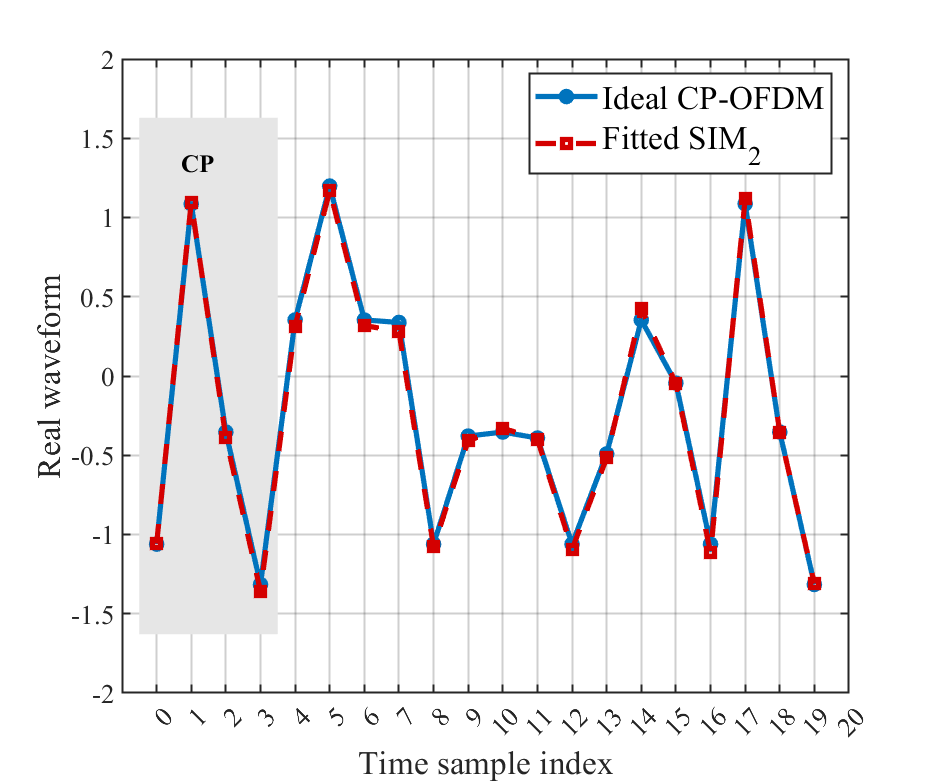}}
    \vspace{-0.2 cm}
\caption{Validation of the proposed wave-domain CP-OFDM materialization by the fitted SIM$_2$.
(a) Frequency-domain responses of several representative SIM$_2$-generated OFDM tones.
(b) CP-extended time-domain waveform generated by the fitted SIM$_2$ compared with the ideal IDFT-and-CP waveform.}
    \label{SIM2OFDM}
\vspace{-0.5 cm}
\end{figure}

Fig.~\ref{SIM2OFDM} evaluates the physical CP-OFDM materialization capability of SIM$_2$ from both frequency-domain and time-domain perspectives. Fig.~\ref{SIM2Oa} shows the frequency-domain responses of representative SIM$_2$-generated OFDM tones, where markers denote the integer DFT sampling points after CP removal and percentages indicate the desired-subcarrier energy preservation ratios. Although OFDM subcarriers overlap in the continuous frequency domain, their orthogonality is determined by the integer DFT sampling locations. It can be observed that the fitted SIM$_2$ preserves the zero-crossing structure at the neighboring subcarrier centers. For the selected tones, more than \(99\%\) of the recovered energy remains on the desired subcarrier, indicating that the offline SIM$_2$ optimization successfully compiles the OFDM basis into the wave domain with negligible residual cross-subcarrier leakage. Fig.~\ref{SIM2Ob} further compares the CP-extended waveform generated by the fitted SIM$_2$ with that of the ideal IDFT-and-CP operator. The two curves closely overlap over both the useful OFDM interval and the CP interval, confirming that SIM$_2$ realizes not only the frequency-domain orthogonality structure but also the time-domain CP-OFDM waveform. Therefore, the fitted SIM$_2$ serves as a physical wave-domain counterpart to the conventional digital IDFT/CP module in the proposed architecture.

\begin{figure}
	\centerline{\includegraphics[width=0.35\textwidth]{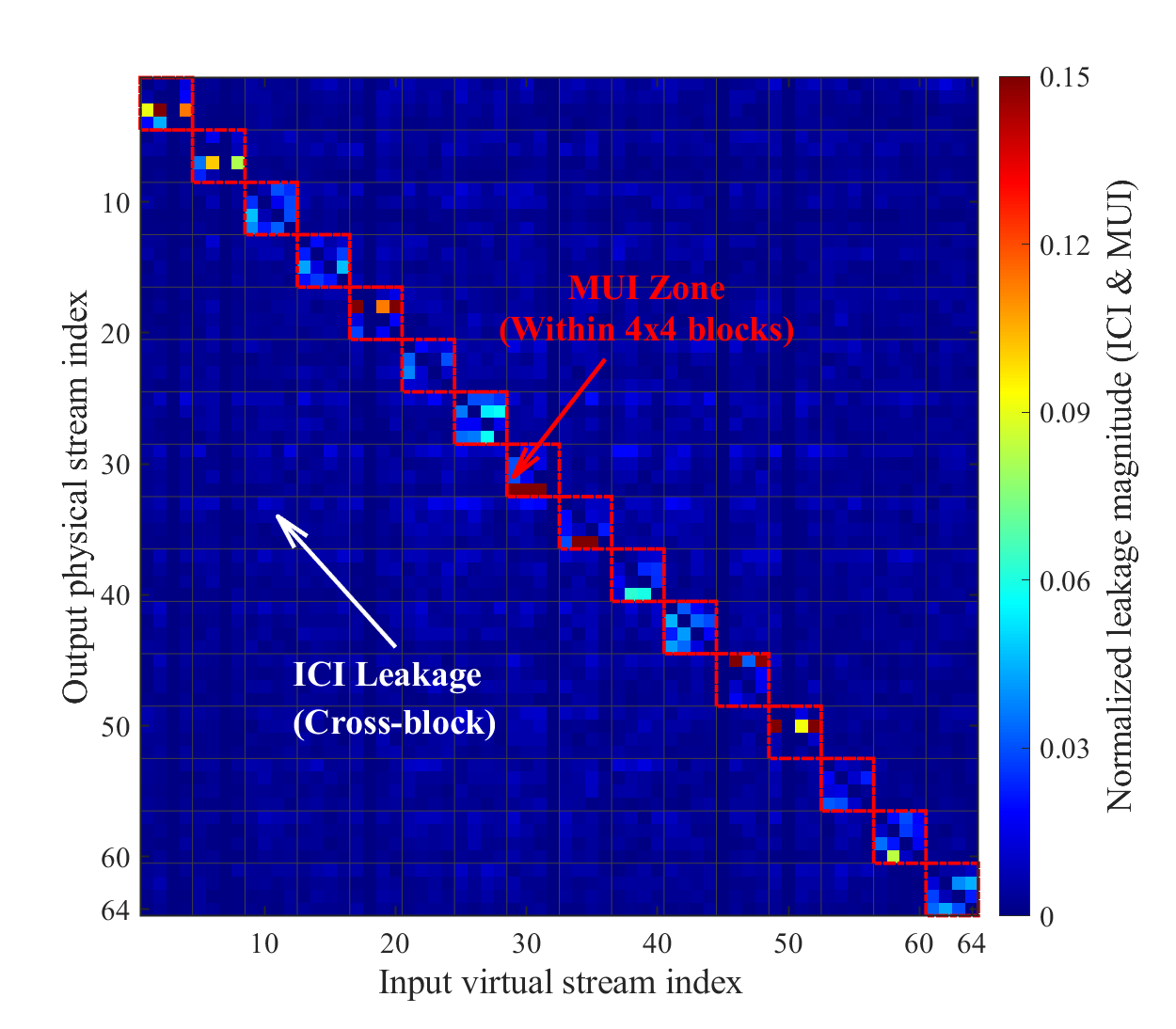}}
\caption{Effective space-frequency leakage map of the cascaded wave-domain transmitter after receiver-side CP removal and DFT. The desired main-diagonal coupling entries are suppressed in the displayed map to expose the residual MUI and ICI.}
	\label{Dia}
\vspace{-0.5 cm}
\end{figure}

To connect this functional validation with the subsequent system-level performance results, Fig.~\ref{Dia} plots the effective space-frequency leakage map after the cascaded SIM transmitter and standard OFDM receiver. For visualization, the desired main diagonal coupling entries are suppressed in the displayed map so that the residual MUI and ICI components can be observed more clearly. Within each $4\times4$ diagonal block, the off-diagonal entries represent residual intra-subcarrier MUI caused by the finite-dimensional, discrete-phase SIM$_1$ precoder. In contrast, the off-block-diagonal entries represent residual ICI induced by the nonideal SIM$_2$ coupling matrix \(\mathbf T^\star\). This visualization highlights an important property of the proposed model: SIM$_1$ imperfections mainly create localized spatial leakage within each subcarrier, whereas SIM$_2$ imperfections create global cross-subcarrier leakage. The following spectral-efficiency results are therefore interpreted through these two residual interference mechanisms.
\vspace{-0.45 cm}
\subsection{Sum Spectral Efficiency versus Transmit Power}
\begin{figure}
	\centerline{\includegraphics[width=0.45\textwidth]{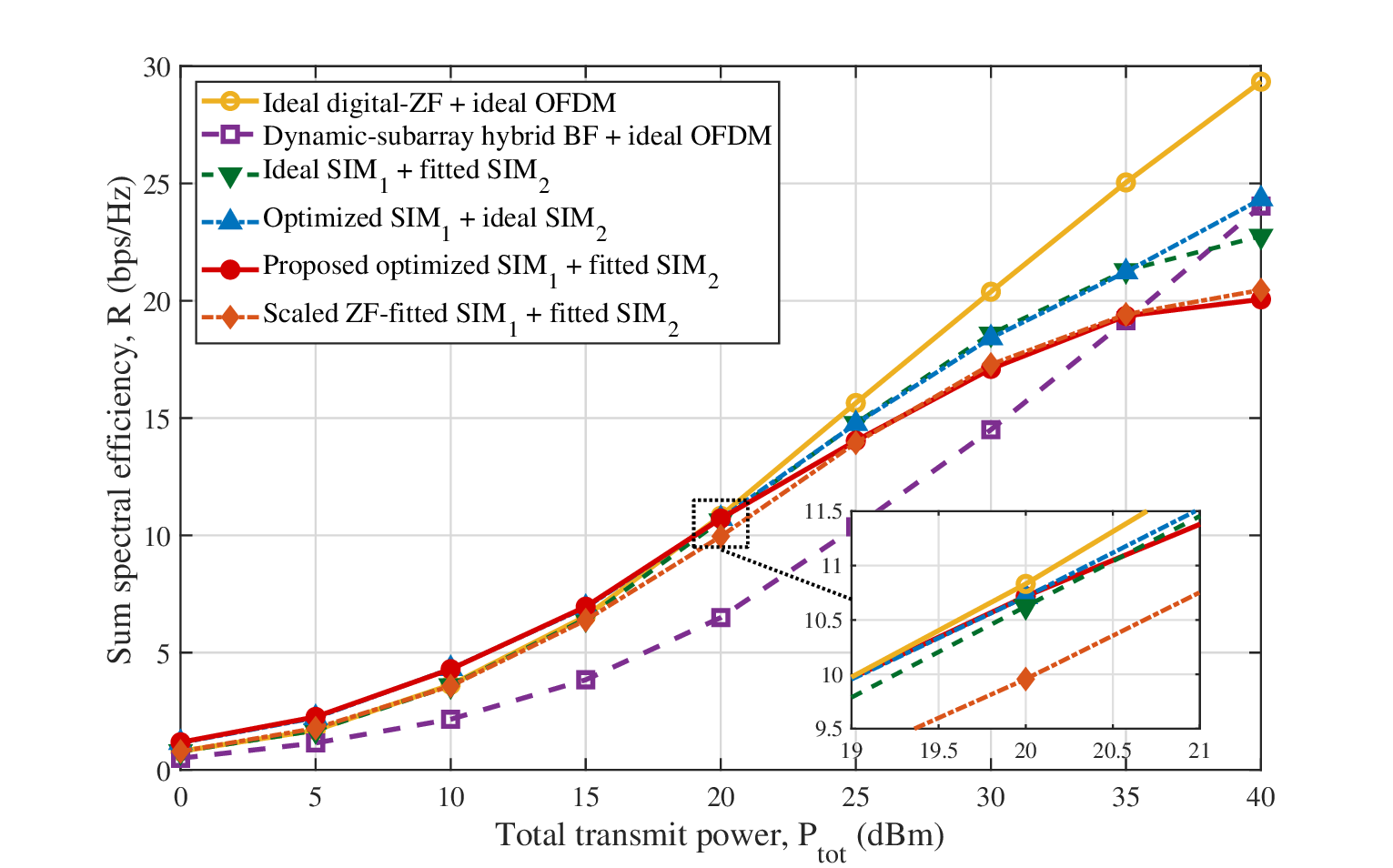}}
\caption{Sum spectral efficiency versus total transmit power for different transmitter architectures.}
	\label{SE1}
\vspace{-0.5 cm}
\end{figure}

Finally, we compare the sum spectral efficiency of different transmitter architectures. Several representative baselines are included: 1) the ideal digital-ZF transmitter with ideal OFDM, which serves as a fully-digital reference; 2) dynamic-subarray hybrid beamforming with ideal OFDM \cite{HY2}, which represents a conventional baseband-assisted hybrid architecture; 3) the cascade of ideal SIM$_1$ and fitted SIM$_2$, which isolates the performance loss caused by SIM$_2$-induced residual ICI; 4) the cascade of optimized SIM$_1$ and ideal SIM$_2$, which isolates the wave-domain spatial precoding loss of SIM$_1$; and 5) the cascade of scaled ZF-fitted SIM$_1$ \cite{Li3} and fitted SIM$_2$, which provides a strong operator-fitting reference. For fairness, the fully-digital ZF and hybrid beamforming baselines use ideal OFDM, the same total power constraint \(P_{\rm tot}\), and the same FP/KKT stream-subcarrier power-loading routine.

Fig.~\ref{SE1} shows that the proposed SE-oriented SIM$_1$ optimization slightly outperforms the digital-ZF benchmark in the low-power regime. By directly maximizing the sum spectral efficiency, the proposed SIM$_1$ optimizer can trade a small amount of residual MUI for stronger desired-signal enhancement and more favorable stream-subcarrier power loading than digital-ZF. As $P_{\rm tot}$ increases, the SIM-based curves gradually become interference-limited due to residual MUI from SIM$_1$ and residual ICI from the fitted SIM$_2$ coupling matrix $\mathbf T^\star$, which explains the high-power saturation trend of the fully wave-domain transmitter. Although the scaled ZF-fitted SIM$_1$ baseline improves the high-power behavior by using auxiliary per-subcarrier scaling factors during digital-ZF fitting, these additional variables introduce extra fitting DoF and are not adopted in the proposed hardware-constrained design. Therefore, the proposed method achieves competitive spectral efficiency with a physically consistent baseband-free transmitter, relying only on discrete SIM phase control and stream-subcarrier power loading instead of conventional digital baseband precoding or auxiliary fitting gains.

\begin{figure}
	\centerline{\includegraphics[width=0.45\textwidth]{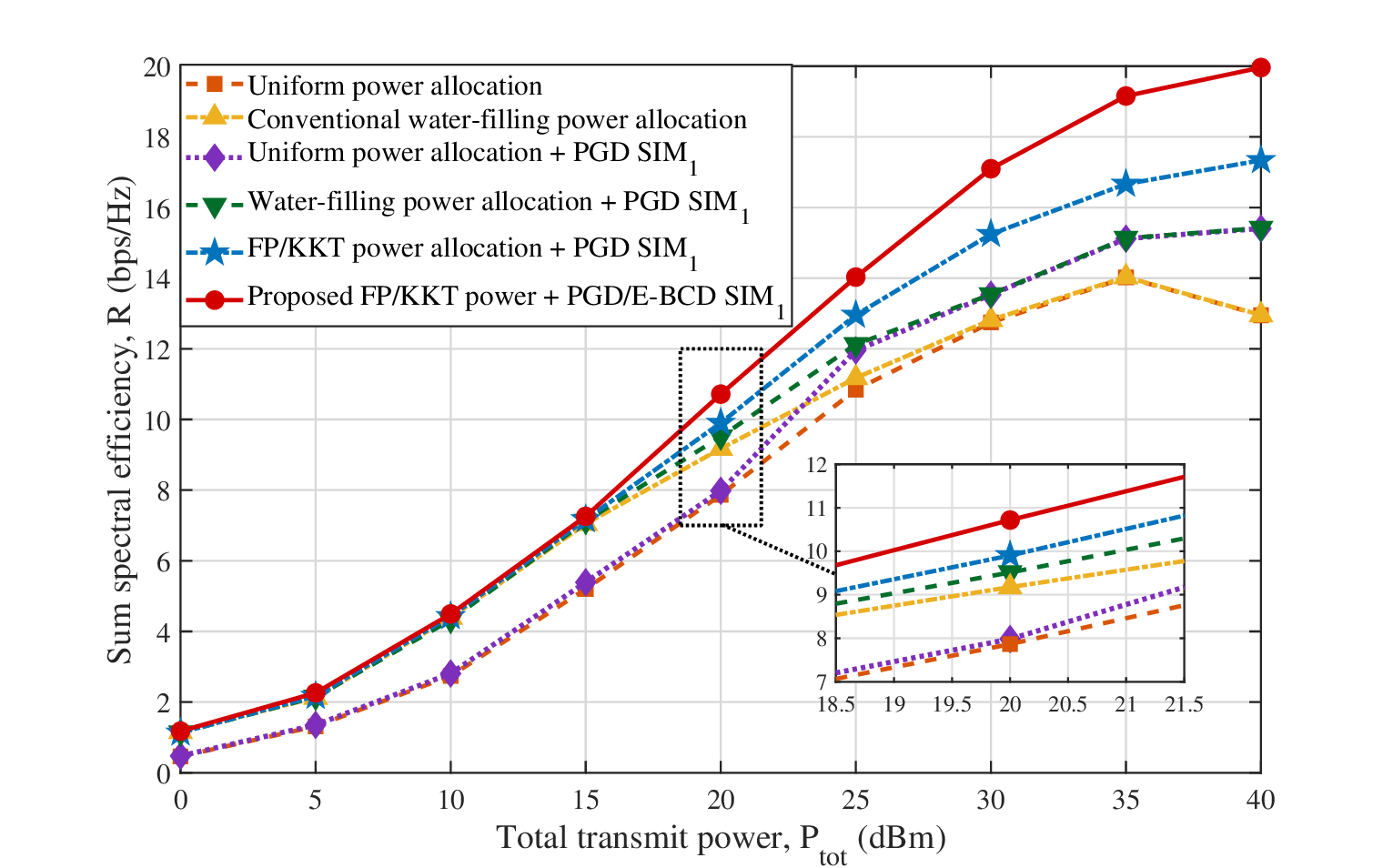}}
\caption{Impact of power allocation and discrete SIM$_1$ optimization on the sum spectral efficiency.}
	\label{SE2}
\vspace{-0.5 cm}
\end{figure}

Fig.~\ref{SE2} evaluates the contributions of the proposed stream-subcarrier power loading and discrete SIM$_1$ phase optimization. The proposed FP/KKT power allocation combined with PGD/E-BCD SIM$_1$ optimization achieves the highest sum spectral efficiency over almost the entire transmit-power range, confirming that both interference-aware power loading and discrete phase refinement are necessary for the proposed fully wave-domain transmitter. Compared with uniform power allocation, the proposed FP/KKT update exploits the heterogeneous stream-subcarrier effective channels and avoids inefficiently spreading power over weak or interference-sensitive dimensions. Compared with conventional water-filling, the proposed power loading is better matched to the coupled SINR expression induced by residual MUI and SIM$_2$-induced ICI. At low transmit powers, the system is mainly noise-limited, and the residual interference terms are weak; hence, conventional water-filling behaves similarly to the proposed FP/KKT allocation and provides comparable performance. However, at high transmit powers, the nonideal fitted SIM$_2$ coupling matrix $\mathbf T^\star$ introduces power-dependent residual ICI, while imperfect SIM$_1$ phase control also leaves residual MUI. As a result, the performance of water-filling can saturate or even decrease in the high-power regime and eventually becomes close to uniform power allocation. In contrast, the proposed FP/KKT update is derived from the actual interference-coupled SINR structure and therefore adapts the power allocation to the residual MUI/ICI pattern. The PGD-only baselines further show that continuous phase refinement alone is insufficient under discrete phase constraints, whereas the E-BCD refinement directly improves the discrete phase codewords and reduces the residual interference floor. These results demonstrate that the proposed FP/KKT power loading and PGD/E-BCD phase optimization jointly contribute to the observed gain.

\section{Conclusion}

This paper has proposed a fully wave-domain wideband MU-MIMO OFDM transmitter enabled by a cascaded SIM structure, which is functionally partitioned into two consecutive blocks for design and optimization. The proposed architecture has physically redistributed the transmitter-side processing chain, including symbol loading, MU-MIMO precoding, IDFT, and CP insertion, from digital baseband to EM wave propagation. SIM$_1$ has realized channel-adaptive symbol loading and MU-MIMO precoding, while SIM$_2$ has materialized the CP-OFDM waveform through an offline IDFT-and-CP operator fitting. This baseband-free architecture has provided an alternative physical realization of the conventional MU-MIMO OFDM transmitter, where spatial precoding and temporal waveform synthesis are mapped to distinct wave-domain modules with different update timescales.

To address the residual ICI introduced by nonideal SIM$_2$-based OFDM materialization, the fitted SIM$_2$ response has been mapped into an effective OFDM-domain coupling matrix and explicitly incorporated into the subsequent SIM$_1$ optimization. Under this nonideal coupling, SIM$_1$ has been optimized according to the end-to-end CP-aware sum spectral efficiency, jointly balancing desired-signal enhancement, residual MUI suppression, residual ICI mitigation, and stream-subcarrier power loading. Simulations have validated the convergence, phase-resolution behavior, architecture trade-off, CP-OFDM materialization accuracy, and sum spectral efficiency performance of the proposed transmitter. Moreover, results have also demonstrated that moderate-resolution phase shifters are effective in the considered setup, increasing SIM depth or width provides diminishing returns, and the remaining performance gap can be interpreted through residual MUI from SIM$_1$ and residual ICI from SIM$_2$. 

More broadly, this work has suggested that a wideband transmitter need not be organized strictly around a digital-baseband-first signal chain. By separating channel-adaptive spatial computation from CSI-independent waveform materialization, the proposed architecture has provided a concrete route for mapping key OFDM transmitter functions into reconfigurable EM structures. Moving from this functional demonstration to practical deployment will require further study of sampling-rate metasurface control, broadband dispersion and loss, phase/amplitude calibration, mutual coupling, and robustness under hardware and CSI uncertainty. These issues motivate future research on prototype-driven EM communication co-design, robust wave-domain optimization, and extensions to multi-stream transmission, adaptive numerologies, coded links, and wave-domain receiver processing.

\end{document}